\documentclass{aa}  

\usepackage{graphicx}
\usepackage{txfonts}
\usepackage[colorlinks=true,linkcolor=blue,citecolor=blue,filecolor=blue,urlcolor=blue]{hyperref}
\usepackage{graphicx}
\usepackage{multirow}
\usepackage{makecell} 
\setcellgapes{5pt}

\begin{document}

   \title{Multiscale entropy analysis of astronomical time series}

   \subtitle{Discovering subclusters of hybrid pulsators}

   \author{J.~Audenaert\inst{\ref{ivs},\ref{MIT}}
          \and
           A. Tkachenko\inst{\ref{ivs}}
          }

   \institute{Institute of Astronomy, KU Leuven,  Celestijnenlaan 200D, 3001, Leuven, Belgium\\
              \email{jeroen.audenaert@kuleuven.be}
    \label{ivs}
    \and
    Department of Physics and Kavli Institute for Astrophysics and Space Research, Massachusetts Institute of Technology, Cambridge, MA 02139, USA
    \label{MIT}
             }

   \date{Received March 3, 2022; accepted Jun 22, 2022}

 
  \abstract
   {The multiscale entropy assesses the complexity of a signal across different timescales. It originates from the biomedical domain and was recently successfully used to characterize light curves as part of a supervised machine learning framework to classify stellar variability.}
   {We aim to explore the behavior of the multiscale entropy in detail by studying its algorithmic properties in a stellar variability context and by linking it with traditional astronomical time series analysis methods and metrics such as the Lomb-Scargle periodogram. We subsequently use the multiscale entropy as the basis for an interpretable clustering framework that can distinguish hybrid pulsators with both p- and g-modes from stars with only p-mode pulsations, such as $\delta$~Scuti ($\delta$~Sct) stars, or from stars with only g-mode pulsations, such as $\gamma$~Doradus ($\gamma$~Dor) stars.
   }
   {We calculate the multiscale entropy for a set of \textit{Kepler} light curves and simulated sine waves. We link the multiscale entropy to the type of stellar variability and to the frequency content of a signal through a correlation analysis and a set of simulations. The dimensionality of the multiscale entropy is reduced to two dimensions and is subsequently used as input to the HDBSCAN density-based clustering algorithm in order to find the hybrid pulsators within sets of $\delta$~Sct and $\gamma$~Dor stars that were observed by \textit{Kepler}.
   }
   {We find that the multiscale entropy is a powerful tool for capturing variability patterns in stellar light curves. The multiscale entropy provides insights into the pulsation structure of a star and reveals how short- and long-term variability interact with each other based on time-domain information only. We also show that the multiscale entropy is correlated to the frequency content of a stellar signal and in particular to the near-core rotation rates of g-mode pulsators. We find that our new clustering framework can successfully identify the hybrid pulsators with both p- and g-modes in sets of $\delta$~Sct and $\gamma$~Dor stars, respectively. The benefit of our clustering framework is that it is unsupervised. It therefore does not require previously labeled data and hence is not biased by previous knowledge.}
   {}


   \maketitle
%

\section{Introduction}

The light curve of a star constitutes a window into its internal and surface structure. By modeling the temporal brightness variations with asteroseismic techniques, we can determine stellar masses and ages, as well as rotation, mixing, and core properties \citep[e.g.,][for a review]{Aerts2021}. In order to determine the stellar properties with the highest precision, stellar variability studies require long and detailed photometric brightness measurements. Space missions such as \textit{Kepler} \citep{borucki2009} and TESS \citep[Transiting Exoplanets Survey Satellite, ][]{ricker2015} have therefore revolutionized the field by delivering continuous months- to years-long high-cadence and high-precision measurements of thousands to millions of stars.

One of the key challenges for large astronomical surveys with high-cadence, long-term light curves such as \textit{Kepler} and TESS lies in identifying the targets of interest. Only after the light curves have been classified according to their stellar variability types can detailed characterizations of the observed stars and planets be made. Given that space missions such as \textit{Kepler} and TESS observe vast amounts of stars, the best ways to achieve this are via crowdsourcing, such as done by \citet{Eisner2021} for TESS planet candidates, or via automated machine learning methods, which is the focus of this paper. \citet{Debosscher2007,Debosscher2009,Debosscher2011}, \citet{Blomme2010,Blomme2011}, and \citet{Sarro2009} laid the foundations for applications of such methods to high-cadence light curves assembled from space. These authors developed dedicated machine learning frameworks to automatically classify space-based light curves according to their stellar variability type. More recently, \citet{Armstrong2016} classified different types of variable stars in the K2 mission fields with self-organizing maps and a random forest classifier. \citet{Hon2018a,Hon2018b,Hon2019}, on the other hand, focused solely on extracting solar-like oscillations in red giants in \textit{Kepler}, K2 and simulated TESS data with convolutional neural networks. \citet{Kuszlewicz2020} went into more detail and classified red giants according to their evolutionary states. \citet{Battley2022} again used self-organizing maps to differentiate the TESS light curves of young eclipsing binaries and transiting objects from other types of variability. The TESS Data for Asteroseismology (T'DA) working group combined multiple separate classifiers into one large classifier to classify TESS lights curves according to their high-level variability types \citep{Audenaert2021}. They validated their classifier on a set of truncated 27.4d Kepler light curves (to mimic single sector TESS data) and applied it to the 167,000 stars observed in Q9. \citet{Barbara2022} also classified \textit{Kepler} Q9 light curves, but used full 90d light curves and specifically focused on the 12,000 A and F stars in the data set, using a classifier based on Gaussian mixtures. \citet{Hon2021} went beyond pure classification and created a full all-sky Gaia-asteroseismology mass map for the red giants they discovered in the TESS data with their convolutional neural network.

Most of the research has focused on using supervised learning to classify light curves \citep[with some exceptions being, e.g.,][]{valenzuela2018, modak2018} or on using unsupervised methods to create a latent space and then subsequently apply (or plan to apply) supervised methods to classify the data based on their positions in the latent space \citep[see, e.g.,][]{Armstrong2016,Battley2022}. Although supervised learning is ideal for structuring large amounts of data, it might be less efficient for smaller data sets where only lower amounts of labeled data are available. Unsupervised classification methods are better suited for this case as they can naturally cluster the data and are not bound by previous knowledge. If used with physically or mathematically interpretable features, their output can also be understood rather easily \citep{molnar2019}.

Here, we explore the use of unsupervised learning to discover subclusters of pulsational variability, given that only limited amounts of data are available in this setting and physical interpretations are important. We specifically use the multiscale entropy (MSE) as the basis for an interpretable clustering framework. The MSE \citep{Costa2002, Costa2005} was introduced to the astronomical domain by \citet{Audenaert2021} and measures the regularity of a time series at different timescales, creating an overall complexity profile. The profile can be used to, for example, distinguish stochastic from deterministic signals and hence, distinguish between different types of stellar variability. This was done by \citet{Audenaert2021}, who used it as one of the features in their stellar variability classification framework. In contrast to Fourier-based methods, such as those based on the Lomb-Scargle periodogram \citep{Lomb1976,Scargle1982}, which provides the frequencies of the modes propagating the stellar interior or the periods of orbiting (sub)stellar companions \citep{Debosscher2009,Blomme2011}, the MSE is a subtler characterization of the light curve that provides insight into the structure of the variability.

We take p- and g-mode pulsators of spectral types F and A as a typical example of variable stars that have partially overlapping instability strips in the Hertzsprung--Russell diagram and give rise to a class of hybrid pulsators with both p- and g-modes \citep[see, e.g.,][]{Uytterhoeven2011,Bowman2018}. Pressure modes, or p-modes, are acoustic waves in the envelope of the star for which the pressure force is the dominant restoring force. In gravity modes, or g-modes, on the other hand, buoyancy is the dominant restoring force. The latter modes mostly probe the deep interior of the star. Our class of p-mode pulsators includes the classical set of $\delta$\,Sct stars, while our class of g-mode pulsators includes, in addition to the classical set of $\gamma$\,Dor stars, stars with Rossby modes. We include Rossby modes under this definition as they are not easily distinguished from the heat-driven g-modes in F-type stars in the time domain and so far Rossby modes are only detected in F-type stars that also reveal g-modes. In Rossby modes, or r-modes, the Coriolis force acts as the dominant restoring force.

$\delta$\,Sct stars are a class of variables that pulsate in radial and non-radial pressure (p) modes \citep{Aerts2010}. Their masses range from 1.5 to 2.5 M$_{\odot}$ and their pulsation periods from about 15 min ($1111.11 \mu$Hz or $96$ d$^{-1}$) to 8 hours ($34.72 \mu$Hz or $3$ d$^{-1}$) \citep{Aerts2010}. $\gamma$\,Dor stars, on the other hand, are a class of variables pulsating in high-radial-gravity (g) modes \citep{Aerts2010}. Asteroseismic modeling revealed the masses of $\gamma$\,Dor stars to range from 1.3 to 1.9 M$_{\odot}$ \citep{Mombarg2019} and typical pulsation periods from 0.3 d ($38.54 \mu$Hz or $3.33$ d$^{-1}$) to 3 d ($3.82 \mu$Hz or $0.33$ d$^{-1}$) \citep{Guzik2000,VanReeth2015b,Li2020}. Rapid rotation can also shift the observed g-mode frequencies toward higher values due to the strong joint influences of the Doppler effect and the Coriolis force. This means that in the lower part of the p-mode frequency region, we might actually also see g-mode pulsations that were shifted upward in frequency due to rapid rotation, as found for both B-type and F-type g-mode pulsators \citep{Aerts2005,Saesen2010,Saesen2013,Mowlavi2013,Mowlavi2016,Mozdzierski2014,Mozdzierski2019,Gebruers2022,deridder2022}.

The instability regions of $\delta$\,Sct and $\gamma$\,Dor stars overlap in the Hertzsprung--Russell diagram \citep{Dupret2004}, giving rise to the class of hybrid pulsators in which the $\delta$\,Sct and $\gamma$\,Dor pulsation excitation mechanisms occur simultaneously \citep{Dupret2005}. Hybrid pulsators thus have both p- and g-modes at the same time \citep[for observational studies, see, e.g.,][]{Handler2002,Balona2011,Uytterhoeven2011,Bradley2015}, where mostly one of the two types of modes is the dominant one \citep{Grigahcene2010}. While p-modes allow us to probe the stellar envelope, g-modes allow us to probe the properties of the near-core region. Hybrid pulsators are interesting targets for asteroseismic studies as they allow for detailed characterizations of stellar rotation profiles \citep{Kurtz2014,Triana2015}, and can give better insights into the mechanisms that drive p- and g-mode pulsations \citep{Dupret2005}. The combination of both modes in one star thus greatly improves the constraints we can put on the overall structure of a star. More details on modern space asteroseismology can for instance be found in \citet{Aerts2021}.

\citet{Audenaert2021} successfully built a supervised classifier to hunt for $\delta$\,Sct and $\gamma$\,Dor stars. However, they only reveal whether it is a potential $\delta$\,Sct or $\gamma$\,Dor star, and not whether it might be a hybrid pulsator (hybrids are assigned to either of the two classes based on their dominant mode). We therefore develop a methodology to automatically differentiate the hybrid pulsators from, respectively, their pure p-mode and pure g-mode pulsators. By using an unsupervised classifier, we avoid the time-consuming task of having to manually label the light curves, as unsupervised methods do not need training sets. Even more importantly, however, is that unsupervised methods are not bound by previous knowledge. We therefore avoid the issue of having to define a strict boundary between the hybrid pulsators and their ``pure'' counterparts. This is crucial because, physically, the difference between $\delta$\,Sct and hybrid $\delta$\,Sct, and $\gamma$\,Dor and hybrid $\gamma$\,Dor is not strict as they have overlapping instability regions in the Hertzsprung--Russell diagram and their pulsation properties are not distinguishable based on their position. The natural order of the data space that an unsupervised classifier provides is therefore ideally suited to interpret this transition.

The aim of this paper is to (i) provide a detailed description of the MSE, (ii) provide a guide for its use and interpretation with regard to astronomical time series, and light curves in particular, and (iii) use the MSE to cluster p-mode pulsators, g-mode pulsators and their hybrid counterparts. We start by explaining the theory behind the MSE in Sect.~\ref{sect:theory}. We then show how the MSE of a light curve should be interpreted, how it provides insights in the underlying dynamics of a star and how it relates to traditional astronomical data analysis methods such as those based on the Lomb-Scargle periodogram in Sect.~\ref{sect:analysis}. The methodology to respectively cluster hybrids from p-mode and from g-mode pulsators is discussed Sect.~\ref{sect:application}. We conclude the paper in Sect.~\ref{sect:conclusions} with a summary and discussion of the results.

\section{Theory}
\label{sect:theory}

The entropy assesses the uncertainty of a system, and thus also the (un)predictability of time series. The entropy was mathematically developed by \citet{Shannon1948} and is commonly used in information theory and biomedicine. The entropy is a measure of the average amount of information required to represent a variable, or put differently, the amount or disorder or randomness present in a system. The Shannon Entropy is defined as 

\begin{equation}
     H(x) = - \sum_{i=1}^{N} p(x_i) \log p(x_i) = - E[\log p(x_i)],
     \label{eq:shannon}
\end{equation}
where $x$ is a discrete random variable, $p(x_i)$ the probability that outcome $i$ of $x$ occurs, and $E$ the expected value operator. In the case of a light curve, $x_i$ are the brightness measurements and $N$ is number of measurements and represents the length of the light curve.

The entropy $H(x)$ is maximal when all outcomes $i$ of $x$ are different and thus also have the same probability. From a frequentist point of view, this means that each event $x_i$ only occurs once. Hence, they are independent from one another. Given that there are no common values in this case, a large amount of information is needed to store the variable $x$, which is equivalent to a high level of uncertainty. One issue that arises when Eq.~(\ref{eq:shannon}) is used to characterize a light curve, is that the number of values $x$ (that is, the flux) can take, $N$, is relatively large compared to the span of the observed values. This biases the calculation of $p(x_i)$ by pushing the probabilities of each $x_i$ to be nearly equal (i.e., $1/N$), even though certain $x_i$ might only differ by a very small value and are visually approximately equal. One way to resolve this issue would be by binning the data or by using a continuous approximation of $H(x)$ such as the differential entropy.

Time series are a special case of data with a set of unique properties that is not present in nonsequential data. A number of entropy metrics were therefore specifically developed to take advantage of these properties and measure the regularity of time-based systems. \citet{pincus1991} proposed the approximate entropy, a regularity statistic to characterize short and noisy time series. It was developed as a practical implementation of the Kolmogorov-Sinai entropy, as the latter tends to decay toward zero for real-world time series. \citet{Richman2000} proposed the sample entropy as a more robust variation of the approximate entropy. The benefit of the sample entropy is that it is less dependent on the length of the time series and that it has a higher consistency over the input parameter range.

The downside of both the approximate entropy and the sample entropy is that they only calculate the entropy for one particular timescale. This can cause the system to be only partially characterized, and, notably in the case of multi-periodic signals, result in a significant loss of information. Multi-periodic signals are especially prevalent in light curve data sets, as stellar variability tends to be active on multiple timescales \citep{Eyer2008}. In order to address this problem, \citet{Costa2002,Costa2005} proposed the multiscale entropy (MSE) to analyze biomedical signals\footnote{The multiscale entropy was originally developed to analyze cardiac interbeat interval time series from ECG recordings.}. Rather than assessing the entropy of a signal at a single timescale, the MSE calculates the sample entropy across multiple timescales through a coarse-graining procedure. This way the MSE assesses the complexity over the full variability spectrum, rather than on just one timescale. 

The sample entropy \citep[$S_E$,][]{Richman2000} measures the regularity of a signal and is defined as

\begin{equation}
    S_E(m,N,r) = \ln \frac{A}{B} = \ln \frac{\sum_{i=1}^{N-m}n_i^m}{\sum_{i=1}^{N-m}n_i^{m+1}},
    \label{eq:SampEn}
\end{equation}
where $m$ is the number of consecutive time steps to take into account, $N$ the total number of data points in the light curve, $r$ the tolerance margin and $n$ the number of vectors close to a template vector $i$ of dimension $m$. A vector $u_i^m$ is close to a vector $u_j^m$ when $d[u_i^m,u_j^m]\leq r$, where $u_i^m = (x_i, ... , x_{i+m-1})$, $d[..]$ is a distance metric, such as the Euclidean distance, and $r$ the tolerance margin for two vectors of $m$ data points to be considered equal. The use of a tolerance margin $r$ allows the sample entropy to be directly applied to continuous or near-continuous (relatively high number of possible states of the measured variable compared to the number of observations) data such as a light curve. It is usually set to $[0.1, 0.2] \times \sigma_{\text{light curve}}$, where $\sigma$ is the standard deviation.

\begin{figure}
\centering
   \includegraphics[width=9cm]{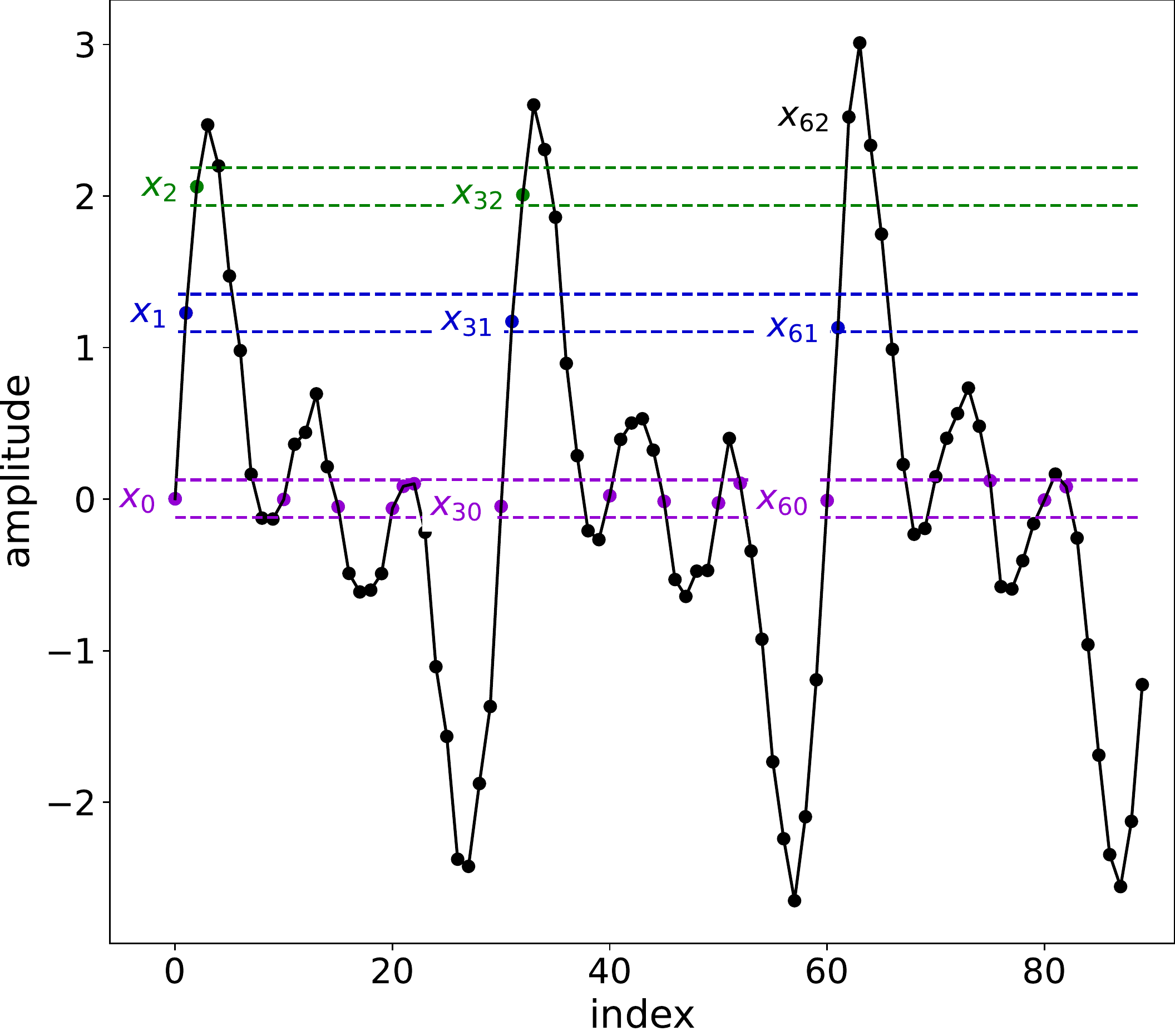}
   \caption{Graphical illustration of the sequence identification procedure that is used to calculate the sample entropy for a simulated light curve. We set $r=0.1 \times \sigma_{\text{light curve}}$ and $m=2$ in this example. The dashed horizontal lines represent $x_0 \pm r$, $x_1 \pm r$ and $x_2 \pm r$. The points that fall within this boundary, namely $d[x_i,x_j] \leq r$, are color-coded in, respectively, purple, blue and green. The first unique sequence or template vector of two components ($m=2$) is $u_{1,m} = (x_0,x_1)$. Two other sequences match this template vector in the light curve, namely the sequences $(x_{30},x_{31})$ and $(x_{60},x_{61})$. Extending $u_1$ to three components ($m+1$) then gives $(x_0,x_1,x_2)$, occurring only once more in the light curve at $(x_{30},x_{31},x_{32})$, as $x_{62}$ does not fall within the tolerance margin $r$ for the extension of $(x_{60},x_{61})$ to $m+1$. There are thus three matches of length $m$ for the template vector $u_1$ and two matches for its extension to length $m+1$. This procedure is then repeated for all other two component template vectors ($u_{i,m+1} = (x_1,x_2)$, ... ,$(x_{N-2},x_{N-1})$) and their three component extensions ($u_{i,m+1} = (x_1,x_2, x_3$), ... , ($x_{N-2},x_{N-1}, x_{N}$)).
    }
    \label{Fig:sampen}
\end{figure}

In order to calculate the sample entropy, we first identify all unique sequences or template vectors of length $m$ that are present in the light curve, where a sequence is thus defined as $u_i^m = (x_i, ... , x_{i+m-1})$. With the term sequence, we thus mean $m$ subsequent data points in the light curve. We then start by counting how many times a certain sequence of length $m$ occurs in the light curve. Next, we extend this unique sequence to length $m+1$ by adding the data point that succeeds the sequence, that is, $x_{i+m}$, and count how many times this extended $m+1$ sequence or pattern occurs in the light curve. This process is then repeated for each unique $m$ and $m+1$ sequence. The tolerance margin $r$ in the sequence identification step allows similar, but not completely equal, sequences to be considered similar. The process is graphically illustrated in Fig.~\ref{Fig:sampen}. The sample entropy is equal to the natural logarithm of the ratio between the sum of the counted $m$ and $m+1$ sequences, A and B, respectively, in Eq.~(\ref{eq:SampEn}). The sample entropy represents the probability that sequences matching each other for the first $m$ components also match for the next $m + 1$ components. Regular or predictable signals, as defined through the recurrence of sequences (or patterns), are assigned lower $S_E$ values while more unpredictable signals are assigned higher $S_E$ values and considered more complex.

The multiscale entropy \citep{Costa2002,Costa2005} extends the sample entropy by calculating it for multiple timescales of the time series. It measures the overall complexity of a light curve, rather than on one scale. The complexity of a signal is in this sense not exactly equal to the entropy of a signal. While the entropy is maximal for a random variable, this is not necessarily the case for the complexity because it is considered fairly easy to quantify randomness. The least complex systems are those with deterministic or predictable signals and those with completely random signals. The most complex systems are those with long-range temporal correlations. In order to account for these long-range correlations, the multiscale approach is required. The different timescales are assessed by coarse-graining the time series, which is essentially equal to running a moving average filter with nonoverlapping windows. Given a time series $(x_i)_{i=1}^N$, coarse-graining is achieved by dividing the time-series into nonoverlapping windows of length $\tau$. Each element $x_j$ of the new time series $(x_j)_{j=1}^{N/\tau}$ is then calculated as

\begin{equation}
    \label{eq:coarse-graining}
    x_{j}^{\tau} = \frac{1}{\tau} \sum_{i=(j-1)\tau + 1}^{j\tau} x_i,
\end{equation}
where $\tau$ is the number of data points in the window (window length), $N$ the number of points in the time series and $j$ the index after coarse-graining.

In order to obtain consistent $S_E$ values, the time series length $N/\tau$ (the length of the series after each coarse-graining step) should at least be somewhere between $10^m$ and $20^m$ \citep{pincus1994}. \citet{Costa2005} showed that the confidence intervals of the $S_E$ values decrease with an increasing number of data points. Given that a single quarter of long-cadence $Kepler$ data consists of more than 4300 data points, we can expect robust results with parameters of $m+1=2$ and $\tau_{max}=10$. This also holds for single sector TESS FFI (30 min cadence) light curves ($\sim 1300$ data points). The MSE will however not be robust for sparsely sampled light curves such as those obtained by Hipparcos \citep{Perryman1997,ESA1997} or Gaia \citep{Gaia2016}.

Entropy metrics are often used in astronomy, but their usage is mostly confined to optimization or modeling problems \citep[see, e.g.,][]{Graham2013,Almeida2020,deFreitas2021}, rather than as features to characterize light curves or stellar variability. 
\citet{Starck2001} also used a multiscale approach to analyze astronomical data, but they used the information at each scale of an image's wavelet transform for the purpose of signal and image filtering, rather than across multiple timescales for the purpose of variability characterization. Applications of the MSE outside of astronomy are manifold and include, for example, assessing whether there is difference between subjects with and without Alzheimer's disease based on Electroencephalography (EEG) signals \citep{Mizuno2010}. The work by \citet{Courtiol2016} is also applied to brain signal analysis, but it is valuable for astronomical applications as well, as they discuss the behavior of the MSE with regard to different theoretical models and provide a set of guidelines for the use and interpretation of the MSE.

\section{Properties from simulations}
\label{sect:analysis}

We empirically analyze the properties of the MSE for a set of already classified \textit{Kepler} long-cadence variable star light curves and for simulated light curves. The results form a set of guidelines for the interpretation and use of the MSE.

\subsection{Effect of coarse-graining}
\label{subsect:coarsegraing}

\begin{figure*}
\centering
   \includegraphics[width=18cm]{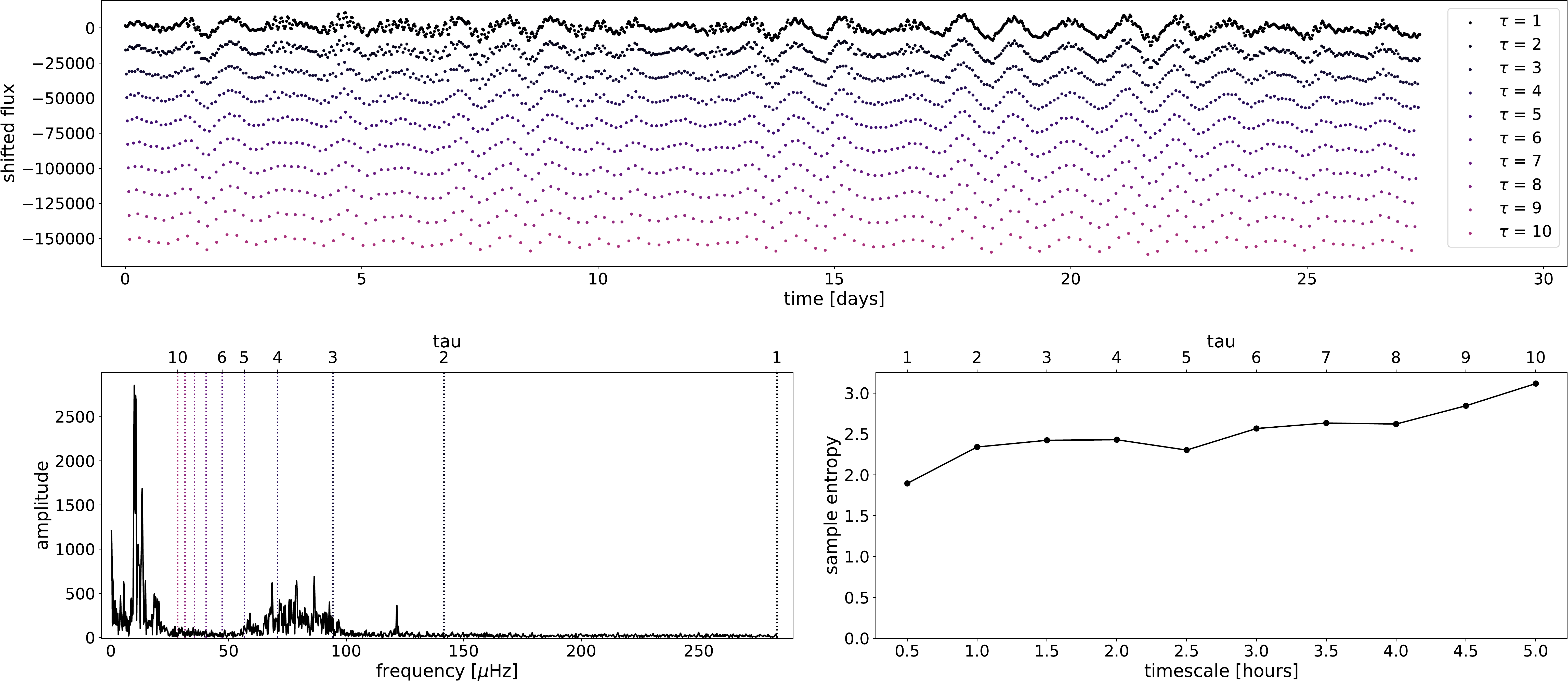}
   \caption{Effect of coarse-graining the $\gamma$\,Dor star KIC005038228. The top panel shows the coarse-grained light curves for an increasing $\tau$, which indicates the scaling factor. We use an offset of 20\,000 per step ($\tau$) for visualization purposes. The bottom left panel shows the amplitude spectrum of the original light curve in black. The dashed lines are the Nyquist frequencies at each scaling factor, and illustrate the low-pass filter effect that occurs due to coarse-graining. It is clear from this that the low-pass filter removes the higher frequencies at each step. The bottom right panel shows the corresponding multiscale entropy curve. Each point on the curve represents the sample entropy calculated for a particular scaling factor.}
    \label{Fig:coarsegraining}
\end{figure*}

We start by demonstrating how coarse-graining a light curve affects its properties. The effect of coarse-graining is two-fold: it acts as a (i) low-pass filter and (ii) downsampler for the light curve \citep{Nikulin2004,Valencia2009}. The first effect, the low-pass filter, removes high-frequency variability from the light curve. This is illustrated in Fig.~\ref{Fig:coarsegraining}, which depicts the light curve, amplitude spectrum and multiscale entropy curve of the $\gamma$\,Dor pulsator KIC005038228. The top panel shows the effect of the scale factor $\tau$ on the variability of the light curve. The light curves are plotted from top to bottom with an increasing $\tau$. It is clear that as $\tau$ increases, the short-term variability is filtered out. The bottom left panel illustrates the high-frequency filtering effect on the amplitude spectrum. The dashed lines indicate the Nyquist frequency at each $\tau$. We see that $f_{\rm Nyquist} = 283.2 \mu \text{Hz}$ for $\tau = 1$, that is, for the original light curve, but that $f_{\rm Nyquist}$ decreases toward 28.3 $\mu\text{Hz}$ for $\tau = 10$, at which it is no longer possible to detect high frequencies. The bottom right panel shows the MSE curve for this star. The points correspond to the sample entropy value of the light curve in the top plot at a particular $\tau$.

\begin{figure*}
\centering
   \includegraphics[width=18cm]{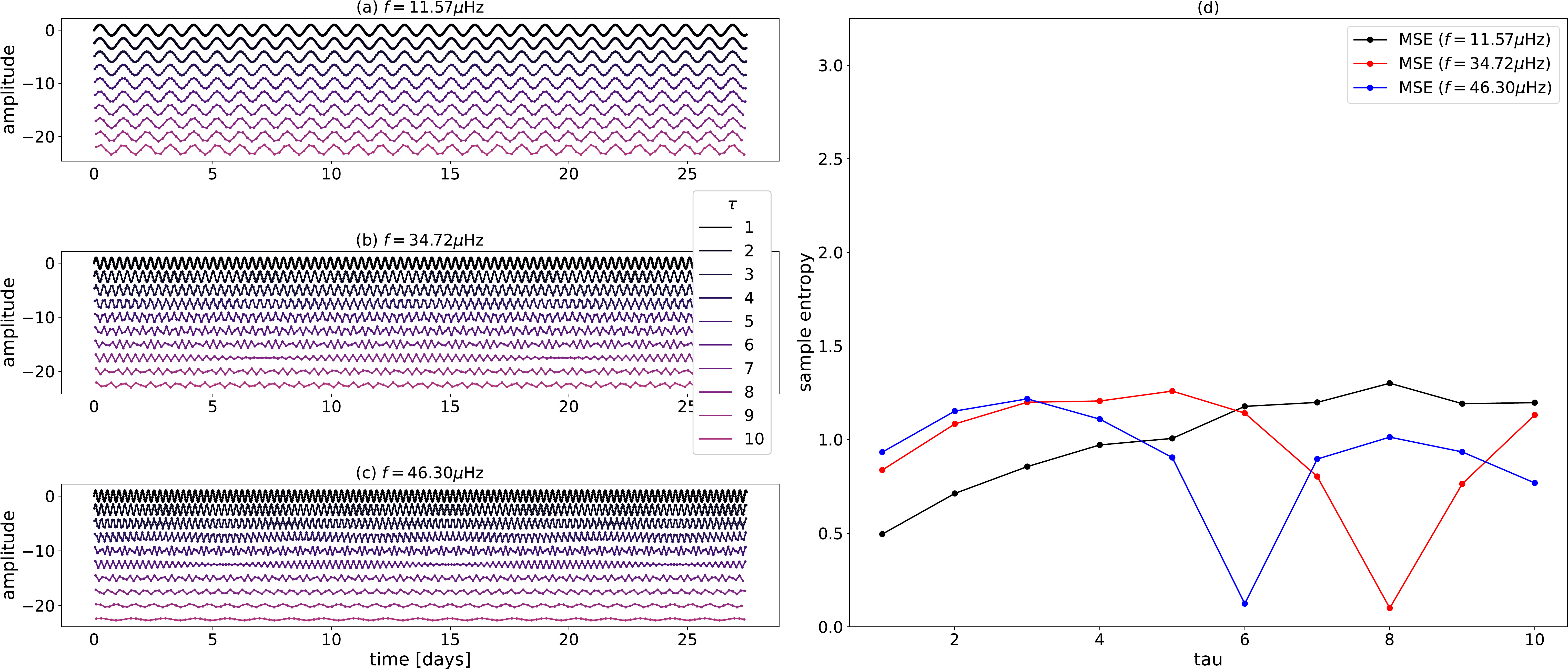}
   \caption{Coarse-grained time series for three different simulated sine waves (left panels; same interpretation as top plot in Fig.\,\ref{Fig:coarsegraining}). The sine waves have an initial frequency of (a) $f = 11.57 \mu \text{Hz}$, (b) $f = 34.72 \mu \text{Hz}$ and (c) $f = 46.30 \mu \text{Hz}$. They are interpreted in the same way as the top panel in Fig.~\ref{Fig:coarsegraining}. The right column (d) shows the MSE curve for each of these different sine waves.}
    \label{Fig:coarsegraining-simulated}
\end{figure*}

The second effect, downsampling, occurs because coarse-graining is achieved by running a moving average filter with nonoverlapping windows. This downsamples the number of points by a factor $\tau$, and hence also the relative level of variability in the light curve. The downside of this method is that, similar to the Fourier domain, aliasing frequencies might be introduced into the coarse-grained signals \citep{Valencia2009}. Aliasing occurs when the Shannon theorem is violated, that is, if the new sampling rate after coarse-graining is less than twice the highest frequency of the signal. The aliasing frequencies usually artifact themselves as sudden drops in the MSE curve. This is especially the case when the dominant frequency is roughly equal to the Nyquist frequency at a particular $\tau$. We demonstrate this with a simulated signal in Fig.~\ref{Fig:coarsegraining-simulated}. The left column displays the coarse-grained signals of three different sine waves, with initial frequencies (a) $f = 11.57 \mu \text{Hz}$, (b) $f = 34.72 \mu \text{Hz}$ and (c) $f = 46.30 \mu \text{Hz}$. We can see that no aliasing occurs in (a) because $f_{\tau=10} < \frac{1}{2T_s}$, where $T_s$ is the sampling period. This requirement does not hold anymore for (b) and (c), causing aliasing frequencies to be introduced by the coarse-graining procedure. The dips for (b) and (c) occur at respectively $\tau=10$ and $\tau=6$, the point at which $f_{\rm Nyquist} \approx f_{\rm signal}$. The aliasing is not directly a problem for the calculation of the MSE though, because the goal of the MSE is to characterize the regularity and complexity of the time series, and not to directly extract the frequencies causing the variability. The sudden drops in the MSE curve that occur due to coarse-graining actually encode the fact that the signal contains relatively high frequencies on this timescale. They are informative with regard to the frequency content of the signal, and thus to the frequencies of the oscillations.

\subsection{Interpretation of MSE curve morphology}
\label{subsect:interpretation}

Understanding the relation between MSE curve morphology and light curve variability is of prime importance if we want to use the MSE to characterize stellar pulsations. We therefore carefully analyze how the shape of a MSE curve should be interpreted.

The amplitude spectrum in the bottom left panel of Fig.~\ref{Fig:coarsegraining} shows the dominant frequencies of the earlier described $\gamma$\,Dor-type star to be in the region around $15\mu$Hz, which is a typical characteristic of g-mode pulsations ($3.8<f<38.5\mu$Hz). There is also a second region with pulsations around $75\mu$Hz, which is typical for p-mode oscillations ($34.7<f<925.9\mu$Hz). The shape of the MSE curve in the bottom right of the figure describes this variability structure by giving an indication of the relative amount of short- and long-term variability that is present in the light curve. When we take a closer look, we see that $S_E$ is lower for $\tau = 1$ than for $\tau = 10$, and that the overall slope is positive. This pattern indicates that there is more long- than short-term variability present in the light curve, which coincides with the hybrid pulsation structure found in the amplitude spectrum. The relatively constant $S_E$ values between $\tau=2$ and $\tau=5$ correspond to the p-modes around $75 \mu\text{Hz}$.

The positive slope for stars with longer periods occurs because the number of $m+1$ sequences (B in Eq.~\ref{eq:SampEn}) decreases more rapidly than the number of $m$ sequences (A in Eq.~\ref{eq:SampEn}). The number of longer sequences decreases faster because only longer term variability patterns remain after coarse-graining, which, due to their nature, occur much less frequently than short-term patterns, and thus result in less matched patterns overall. We see the opposite effect for stars with only short period variability, as the removal of the short period variability makes the curve more constant and more rapidly increases the number of $m+1$ matches given that the whole signal is more similar now.

\begin{figure*}
\centering
   \includegraphics[width=18cm]{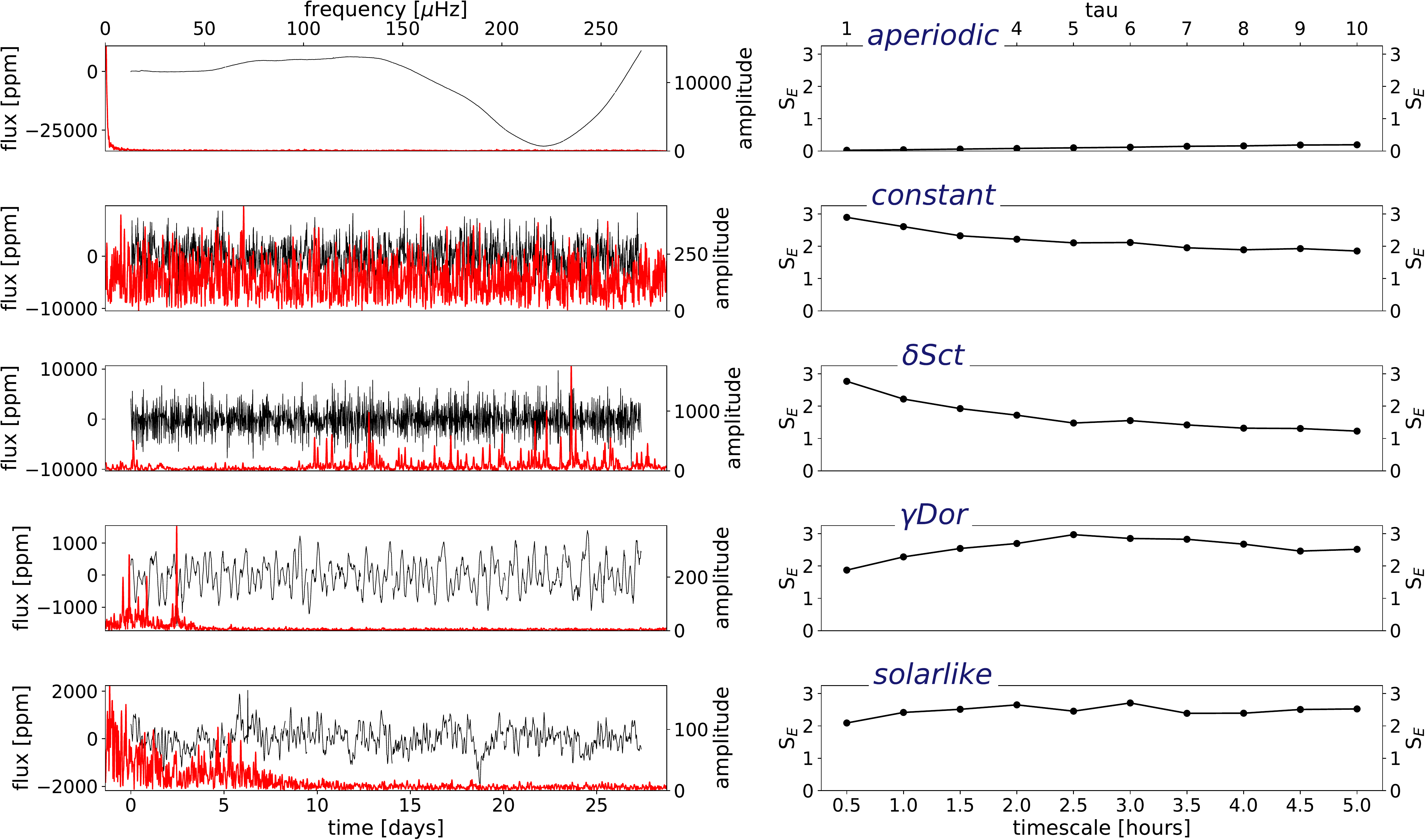}
   \caption{Light curve (left column; black), amplitude spectrum (left column; red), and multiscale entropy curve (right column) for a typical aperiodic, constant, $\delta$\,Sct, $\gamma$\,Dor, and solar-like star.}
    \label{Fig:mse-curves}
\end{figure*}

We confirm this interpretation by comparing the light curves and amplitude spectra for five different types of variability against their MSE curves. The plots are displayed in Fig.~\ref{Fig:mse-curves}. Next to the $\delta$\,Sct and $\gamma$\,Dor classes, we also examine stars of the aperiodic, constant and solar-like pulsator types. Aperiodic stars have light curves that do not show any clear periodic signal on the examined timescale covered by the light curve (e.g., long-period variables), constant stars have light curves without any statistically significant periodic or aperiodic variability, that is, they only contain white noise, while the light curves of solar-like pulsators are dominated by granulation and stochastically excited high-frequency oscillations \citep{Chaplin2013,Garcia2019}. We plot these additional variability types for two reasons. Firstly, they are ideal variability types to benchmark the MSE curves of the $\delta$\,Sct and $\gamma$\,Dor stars against. The light curves of constant and aperiodic stars are theoretical opposites of those of pulsators given that they respectively show no significant activity and no periodicity, while the light curves of solar-like stars have granulation and oscillation patterns that are active on similar timescales. Secondly, these additional variability types could potentially also contaminate automatically selected samples of $\delta$\,Sct and $\gamma$\,Dor stars. This could, in the case of constant stars, occur because their noise is, similar to the oscillations in $\delta$\,Sct stars, active on shorter timescales. The light curves of aperiodic stars, on the other hand, might show low-frequency peaks in the Fourier domain, while the red giants in the solar-like pulsator class have lower frequencies that are in the same region as the $\delta$\,Sct and $\gamma$\,Dor stars.

The plots in Fig.\,\ref{Fig:mse-curves} show that the MSE curves of the $\delta$\,Sct and constant stars both have an opposite slope from that of a $\gamma$\,Dor star. The downward slope of the $\delta$\,Sct star and constant star occur because, respectively, their oscillations take place on short timescales and their noise is only present on short timescales, while the positive slope for the $\gamma$\,Dor star occurs because its g-mode oscillations take place on longer timescales. A negative slope thus indicates that a light curve has more short- than long-term variability, while a positive slope indicates that there is more long- than short-term variability present in the light curve. Although the multiscale entropy curves of the $\delta$\,Sct and constant stars can look similar at first sight, the ensemble analysis with multiple stars in Fig.\,\ref{Fig:mse-ensemble} shows that the MSE curves of constant stars have a much smaller variance and slope compared to those of $\delta$\,Sct stars. This makes sense given that the light curves of constant stars are pure white noise time series coming from the same Gaussian distribution while the $\delta$\,Sct light curves come from actual $\delta$\,Sct stars with pulsations covering a broad frequency region. The sawtooth-like pattern that is seen for some of the $\delta$\,Sct stars is caused by aliasing frequencies, as discussed in the last paragraph of Sect.~\ref{subsect:coarsegraing} and illustrated in Fig.\,\ref{Fig:coarsegraining-simulated}. The MSE curves of aperiodic stars are monotonically increasing because their light curves only contain information on the longest timescales. The MSE curves of solar-like pulsators follow a pattern similar to those of $\gamma$\,Dor stars, but have a smaller slope in the beginning and level off near the end. When coupling this to their physical characteristics, we see that this occurs because the granulation is active at lower frequencies while the oscillations take place at higher frequencies, creating two slightly dispersed active regions in the frequency plot. Their variability is thus more, but not perfectly, balanced between the short- and long-term, hence creating a more balanced MSE curve. Fig.~\ref{Fig:mse-ensemble} demonstrates the MSE's capability to differentiate these five different types of stellar variability.

\citet{Audenaert2021} also showed the capability of the MSE to distinguish pulsating stars from binary systems and rotational variables. Eclipsing binaries could in most cases also be distinguished from contact binaries and from rotational variables. Contact binaries and rotational variables cannot be distinguished from each other because their light curves are nearly identical in the time domain. \citet{Audenaert2021} therefore grouped these two classes into one overarching class. Theoretically, it should also be possible to distinguish pulsators with signals of rotational modulation from pure pulsators. However, more detailed studies are still required to confirm this, as (i) \citet{Audenaert2021} used the MSE in combination with other attributes and (ii) light curves with instrumental trends ended up in the contact and rotational variables class due to the deliberate absence of a separate class.

\subsection{Relation with periodicity}

\begin{figure*}
\centering
   \includegraphics[width=18cm]{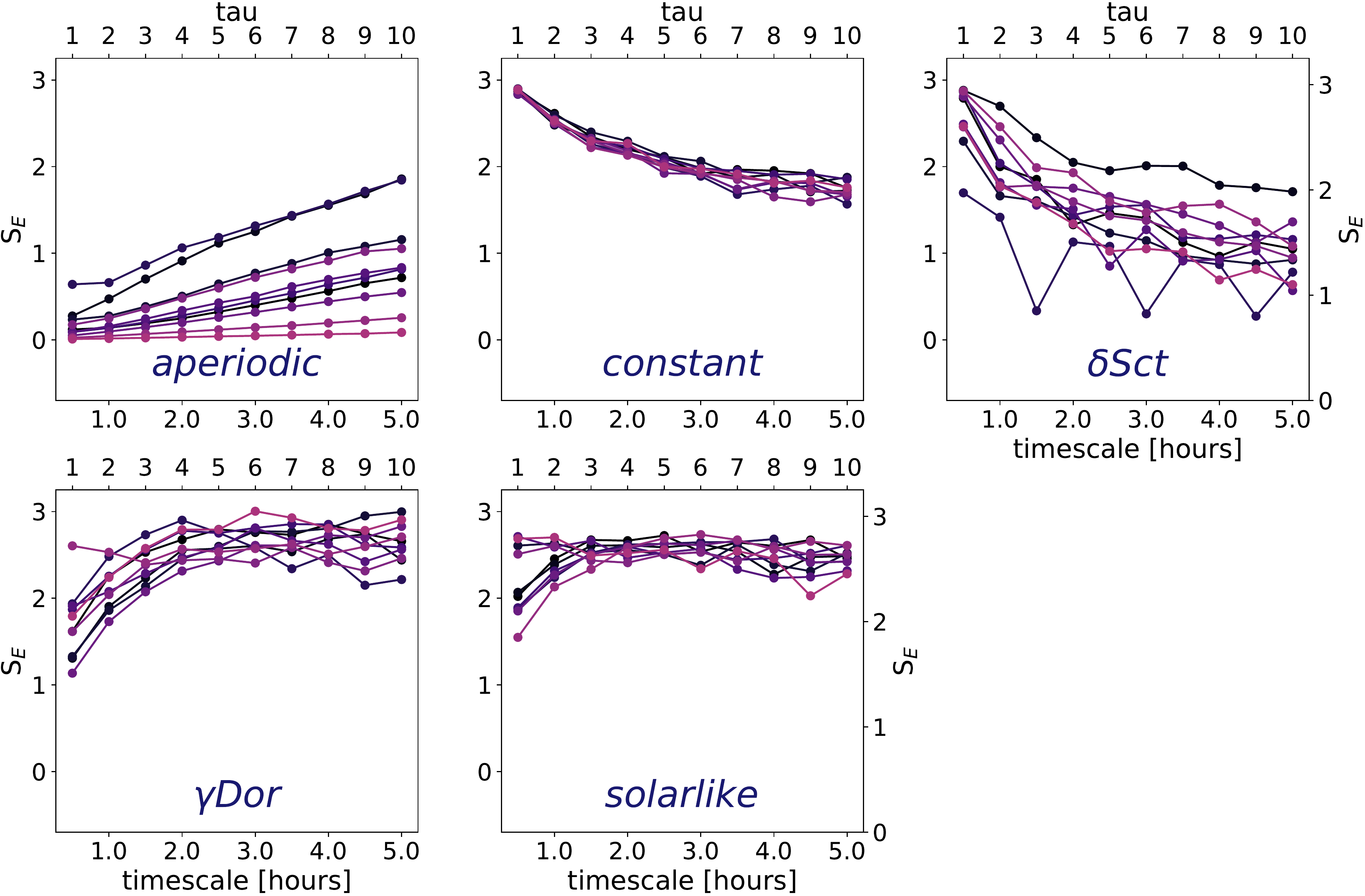}
   \caption{Multiscale entropy curves for ten random stars of the aperiodic, constant, $\delta$\,Sct, $\gamma$\,Dor and solar-like variability type from the training set in \citet{Audenaert2021}.}
    \label{Fig:mse-ensemble}
\end{figure*}

\begin{figure}
\centering
   \includegraphics[width=9cm]{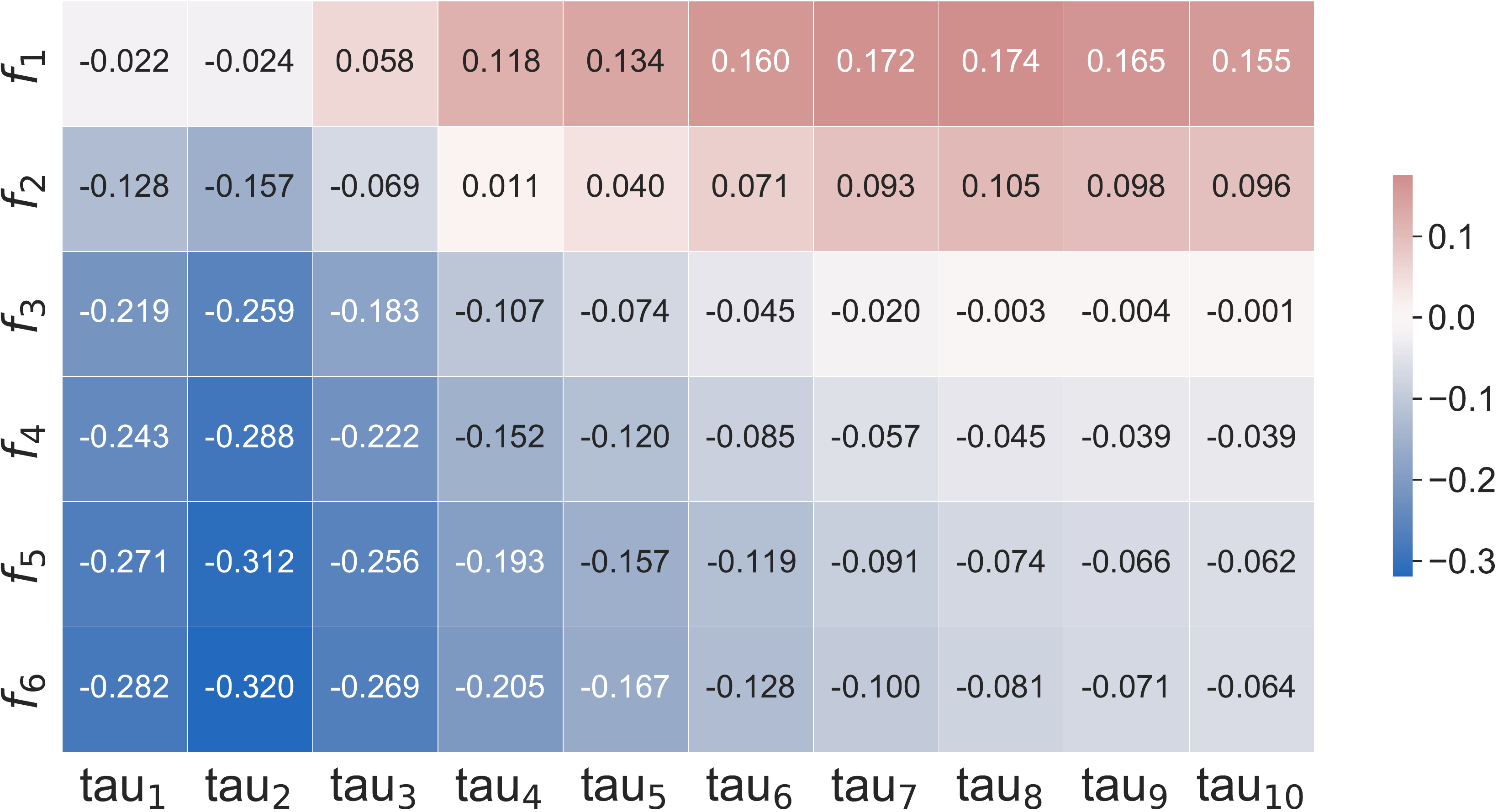}
   \caption{Spearman's rank-order correlation matrix assessing the correlation between the first six significant frequencies (according to the amplitude's signal-to-noise ratio) obtained with Lomb-Scargle and the ten sample entropy values that constitute an MSE curve, calculated based on the values of 8328 stars coming from different variability classes.}
    \label{Fig:correlation-matrix}
\end{figure}

The previous sections showed that, in line with \cite{Bruce2009}, \citet{McIntosh2014}, \citet{Mizuno2010} and \citet{Courtiol2016}, the sample entropy ($S_E$) and multiscale entropy contain information with regard to the frequency content of a signal, or in our case, light curve. We formally test this hypothesis by calculating the Spearman rank correlation matrix between the ten $S_E$ that constitute a MSE curve and the first six significant frequencies obtained with the Lomb-Scargle periodogram. We calculate the matrix based on the amplitude spectra and MSE curves of the 8328 stars that constitute the training set in \citet{Audenaert2021}. The stars in Figs.\,\ref{Fig:mse-curves} and \ref{Fig:mse-ensemble} also come from this training set. The frequencies that were not significant were replaced with -1 in all tests. The correlation matrix is shown in Fig.~\ref{Fig:correlation-matrix}. We see a clear correlation between the sample entropy values at each $\tau$ and the frequencies extracted from the Fourier domain, confirming the relation between the MSE and frequency values.

The second way we test whether the MSE is related to the frequency content of a star, is by training a random forest \citep{Breiman2001} Regressor with the sample entropy values as input and the frequencies as output. This results in a model with $R^2 \approx 0.62$ when the first three frequencies are predicted as output, and $R^2 \approx 0.51$ when the first six frequencies are predicted (according to the amplitude's signal-to-noise ratio), where $R^2$ is the proportion of the variance of the output variables that is explained by the model based on its input variables. A Canonical Correlation Analysis gives similar results. This result was expected from a mathematical point of view, as we showed that a violation of the Shannon theorem results in frequency information being introduced into the MSE curve, and that the slope contains information with regard to the type of variability. Stars with high-frequency oscillations, such as $\delta$\,Sct stars, only contain information on the smallest timescales. Their initial light curves are therefore characterized by a high amount of complexity, but this complexity decreases with an increasing timescale ($\tau$), as the light curve now starts to approach a flatter signal. Stars with low-frequency oscillations on the other hand, such as $\gamma$\,Dor stars, only contain information at larger timescales. Their light curves are therefore characterized by a lower amount of initial complexity, but as the scale increases, more of the signal starts to be taken into account and the complexity also increases. Hence, this relation between the frequencies and MSE.

\subsection{Effect of longer time series}

\citet{pincus1994} already noted that the number of data points affects the calculation of the sample entropy. \citet[][Fig.\,14]{Costa2005} also showed that the confidence intervals of the $S_E$ (see Eq.\,\ref{eq:SampEn}) decrease with the number of data points. The number of data points is an important aspect to be taken into account in high-cadence uninterrupted astronomical data sets, as the number of brightness observations for a star varies per space mission and is often also dependent on its location in the field of view. Depending on the type of oscillations we are interested in, this is of importance to a greater or lesser extent. In order to find stars with low-frequency, low-amplitude g-mode pulsations, for example, we typically need light curves with a longer time base as fewer pulsation cycles are covered in the same period compared to stars with higher-frequency p-modes.

We therefore explore the effect of moving from a time base of 27.4 d to 94 d, 1 yr, 2 yr and 4 yr. These are, respectively, the lengths of single sector TESS data, single Quarter of \textit{Kepler} data, TESS Continuous Viewing Zone (CVZ) data, PLATO \citep{Rauer2014} Long Pointing Field (LPF) data and the full length of the \textit{Kepler} mission. We perform our experiments with long-cadence (30-min) \textit{Kepler} data, as these light curves go up to lengths of 4yr and can easily be truncated to shorter time bases.

\begin{figure}
\centering
   \includegraphics[width=8cm]{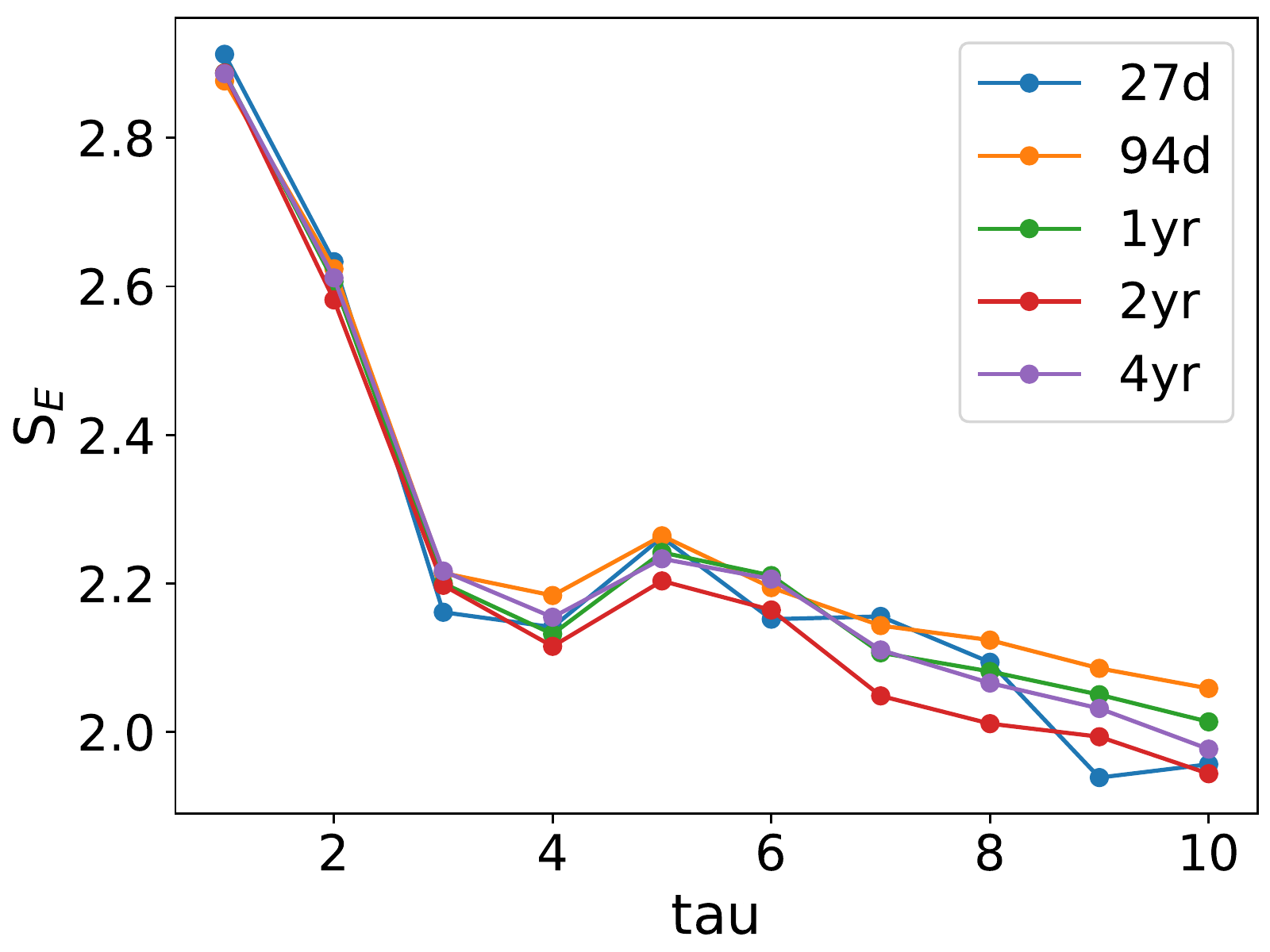}
   \caption{Multiscale entropy curves for the light curve of the $\delta$\,Sct star KIC009655438 truncated at different lengths.}
    \label{Fig:mse-length-dsct}
\end{figure}

\begin{figure}
\centering
   \includegraphics[width=8cm]{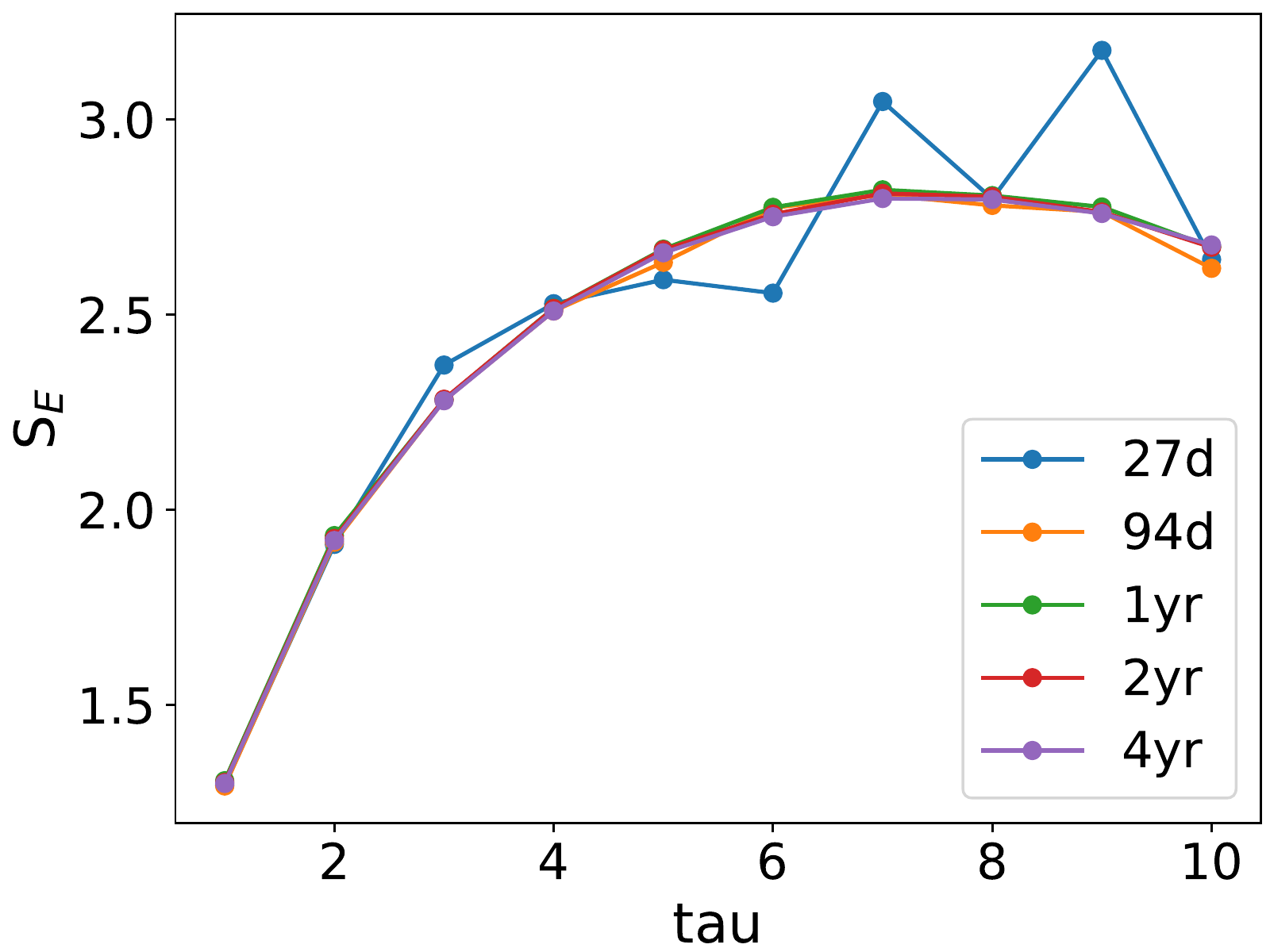}
   \caption{Multiscale entropy curves for the light curve of the $\gamma$\,Dor star KIC003343854 truncated at different lengths.}
    \label{Fig:mse-length-gdor}
\end{figure}

We find that longer data sets have a positive effect on the robustness of the $S_E$ values, but that, starting from a minimum length of $\sim 1000$ data points, the additional increase in robustness is small. The exact effect however, depends on the type of variability (short- vs. long-term). In Fig.\,\ref{Fig:mse-length-dsct}, we show the MSE curves for the $\delta$\,Sct-type star KIC009655438, for which the light curves have been truncated to the lengths described in the previous paragraph. From a time base of 94d ($\sim 4500$ data points) on, the shape of the MSE curve becomes very stable and, apart from a slight downward shift, possibly due to the lower uncertainty that originates from the higher number of data points, the curves do not change significantly. In Fig.\,\ref{Fig:mse-length-gdor}, we show the same plot but for the $\gamma$\,Dor star KIC003343854. We clearly see a larger deviation for the 27 d MSE curve compared to the $\delta$\,Sct case. This is understandable given that $\gamma$\,Dor stars are low-frequency pulsators; hence, fewer pulsation cycles are covered in the same 27 d period. However, overall, increasing the time base does not drastically improve the MSE curve and good results are already obtain for 27 d light curves, although 94 d might be better for low-frequency variability. This is important given that TESS observes millions of stars at a 27.4 d time base. Lowering the number of data points below $\sim 10^m = 1000$, does have a much more negative effect on the MSE curve, as stated in \citet{pincus1994}. We conclude that the MSE applied to light curves of only one month is worthwhile for variability classification.

\section{Discovering hybrids with MSE clustering}
\label{sect:application}

We have shown that the MSE is a powerful tool to characterize the light curve structure of a variable star. It is therefore ideally suited as the basis for a clustering framework that can differentiate between different types of pulsators. We specifically focus on separating hybrid pulsators from pure p- and g-mode pulsators in sets of $\delta$\,Sct and $\gamma$\,Dor stars, as there is no automated tool yet for this that only relies on time domain information, while hybrid stars are prime targets for asteroseismic analysis \citep{Aerts2021}.

\begin{figure*}
\centering
   \includegraphics[width=18cm]{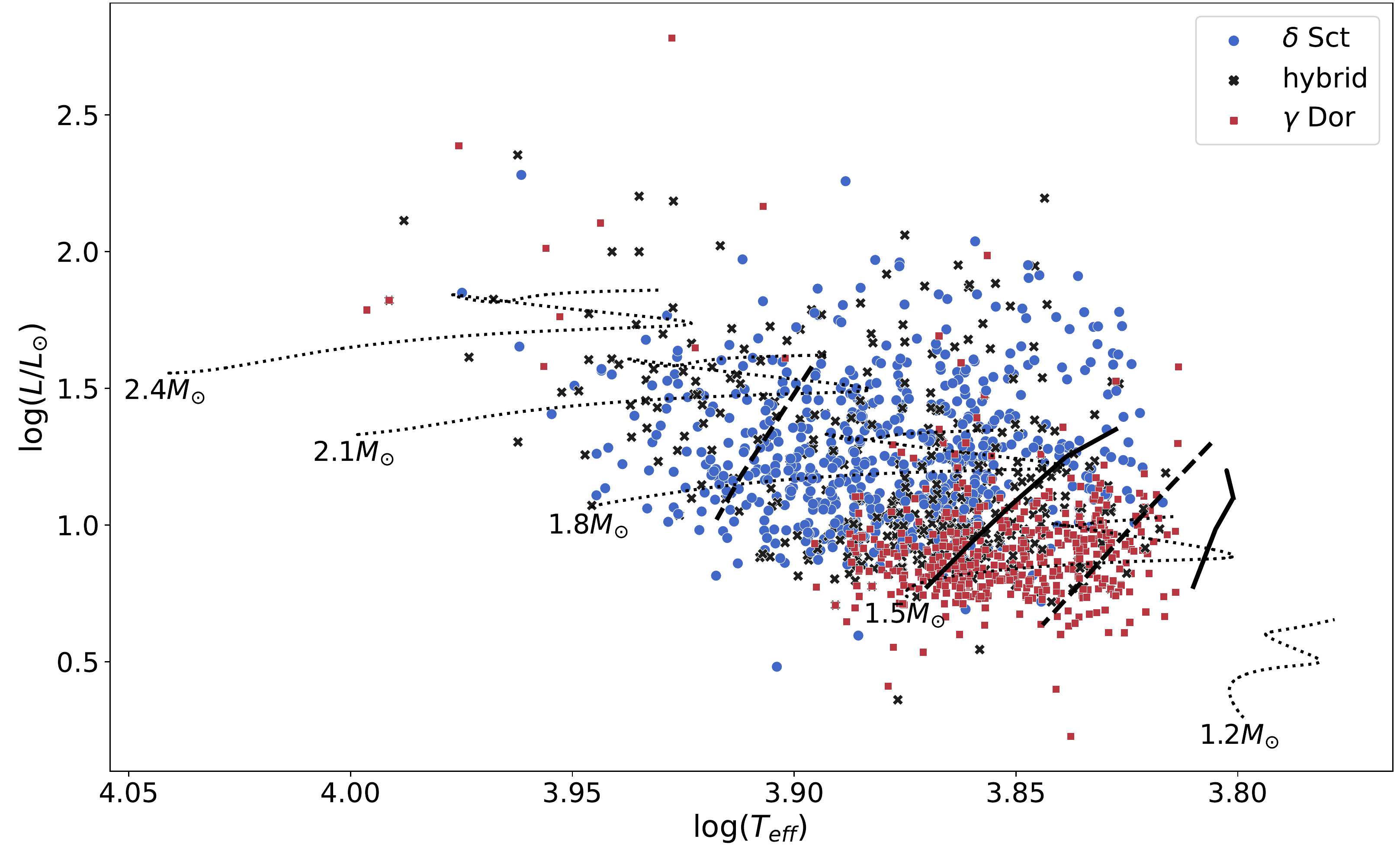}
   \caption{Hertzsprung-Russell diagram showing the log(T$_{\text{eff}}$) and log(L/L$_{\odot}$) values for our sample of $\delta$\,Sct, $\gamma$\,Dor and hybrid pulsators originally from \citet{Bowman2016} and \citet{Li2020}. The colors and marker style indicate the variability class as listed by \citet{Bowman2016} and \citet{Li2020} based on an inspection of the frequencies in the amplitude spectrum. The hybrid stars of both catalogs have been merged as the catalogs do not indicate whether the hybrid stars are dominantly of the $\delta$\,Sct- or $\gamma$\,Dor-type. The black solid and dashed lines respectively indicate the $\delta$\,Sct and $\gamma$\,Dor theoretical instability strips given by \citet{Dupret2005}. The dotted lines are theoretical evolutionary tracks computed by \citet{Johnston2019} for stars between $1.2 M_{\odot}$ - $2.4 M_{\odot}$ and solar metallicity.}
    \label{Fig:hrd}
\end{figure*}

We use the catalog with $\delta$\,Sct stars from \citet{Bowman2016} and the catalog with $\gamma$\,Dor stars from \citet{Li2020} to demonstrate and validate our methodology. We plot the Hertzsprung-Russell diagram with the log(T$_{\text{eff}}$) and log(L/L$_{\odot}$) values for all stars in both catalogs in Fig.\,\ref{Fig:hrd}. We used the effective temperature values from the \textit{Kepler} DR25 input catalog \citep{Mathur2017} and the luminosities from \citet{Murphy2019}. The black solid and dashed lines in the plot respectively show the $\delta$\,Sct and $\gamma$\,Dor instability strips from \citet{Dupret2005}. The strips only give an indication of the locations of the stars in the diagram however, as they were calculated for specific input physics. Therefore, not all stars in our sample fall within these boundaries. The black dotted lines show the theoretical evolutionary tracks computed by \citet{Johnston2019} for stars between $1.2 M_{\odot}$ - $2.4 M_{\odot}$ and solar metallicity, which show that the data are in line with the $\delta$\,Sct and $\gamma$\,Dor star mass ranges. We left out $\sim 99$ stars from these catalogs for which there were no luminosity values reported in \citet{Murphy2019}. We also show a prototypical example of the light curve, amplitude spectrum and MSE curve for a $\delta$\,Sct, $\delta$\,Sct-hybrid, $\gamma$\,Dor and $\gamma$\,Dor-hybrid star in Fig.~\ref{Fig:mse-hybrids}.

\subsection{Methodology}
\label{Subsect:method}

We first computed the multiscale entropy curve from four-year stitched \textit{Kepler} light curves with $\tau_{max}=10$ and $m+1=2$. We then took the 10 $S_E$ values per star that form the MSE\footnote{We used pyEntropy (\url{https://github.com/nikdon/pyEntropy}) for the calculations.} and used  UMAP\footnote{\url{https://github.com/lmcinnes/umap}} \citep[Uniform Manifold Approximation and Projection for Dimension Reduction; ][]{McInnes2018,mcinnes2018umap-software} to produce a two-dimensional equivalent of the ten-dimensional MSE, in order to avoid working in a high-dimensional data space. UMAP is a dimensionality reduction techique that relies on Riemannian geometry and algebraic topology. The output is similar to the t-SNE dimensionality reduction technique \citep{vanderMaaten2008}, but UMAP preserves the global structure of the
data better and has a lower computational complexity. In order to optimally retain the local density structure of the data, we used densMAP \citep{Narayan2021}, UMAP's density-preserving equivalent. densMAP uses an estimate of the local density in the original data space as a regularization parameter in the calculation of the two-dimensional UMAP representation. We specifically used the density-preserving version of UMAP because we cluster the data with a density-based algorithm. The two UMAP components are given as input to HDBSCAN\footnote{\url{https://hdbscan.readthedocs.io/}} \citep[Hierarchical Density-Based Spatial Clustering of Applications with Noise; ][]{McInnes2017, Campello2013}, which then provides a cluster structure of the data space that can be analyzed to find subgroups of stars.

\subsection{\texorpdfstring{$\delta$}{delta} Sct star catalog}
\label{Subsect:dsct}

\begin{figure}
\centering
   \includegraphics[width=9cm]{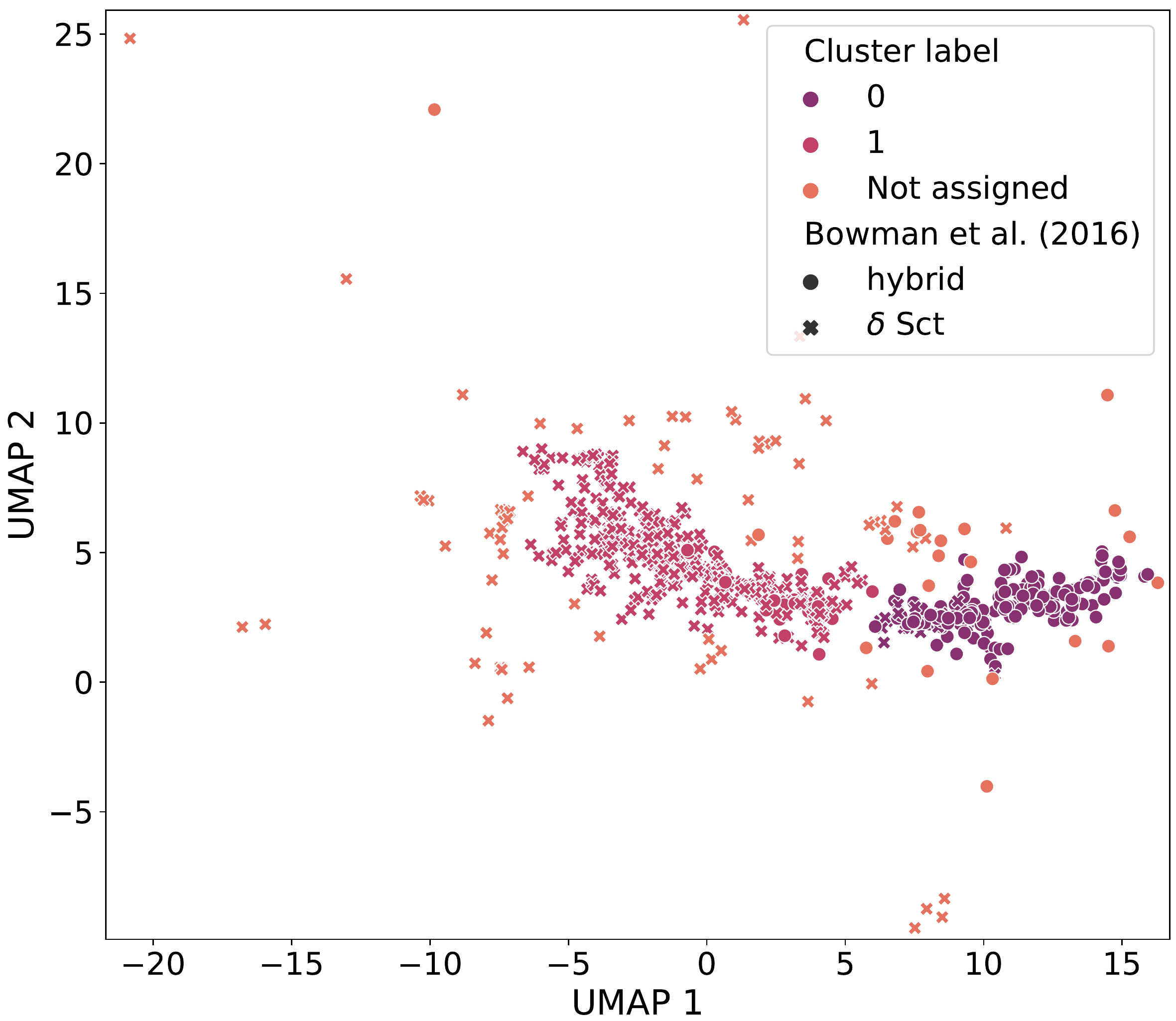}
   \caption{Clustering structure of the $\delta$\,Sct sample from \citet{Bowman2016}. The cluster labels are indicated in color and represent the cluster to which the stars is assigned by HDBSCAN based on the density structure of the UMAP components. The ``Not assigned'' label indicates that the star is seen as noise because it is not close enough to any of the other clusters. The labels from \citet{Bowman2016} are indicated by the marker shapes. Their labels were assigned based on a visual inspection of the light curves and power spectra. These labels were not used during our clustering procedure as it is fully unsupervised. The labels only serve as an indication to better understand the clustering results. The corresponding confusion matrix is shown in Table~\ref{Tab:confusion-dsct}.}
    \label{Fig:clustering-dsct}
\end{figure}

We take the catalog of 983 $\delta$\,Sct stars compiled by \citet{Bowman2016} and use their original stitched 4yr long-cadence \textit{Kepler} PDC\_SAP light curves. We then apply the steps described in Sect.~\ref{Subsect:method} to separate pure $\delta$\,Sct stars with only p-mode pulsations from hybrids with both p- and g-mode pulsations. We set the UMAP parameters to \verb"dens_lambda" = 5.0, \verb"n_neigbours" = 25 and \verb"min_dist" = 0.0, based on UMAP developer advice and our own tests. These parameters respectively control how much of the global and local structure are preserved and how tightly the points are compressed together. For HDBSCAN, we set \verb"min_cluster_size" = 50 and \verb"min_samples" = 5, which respectively control the minimum size of a potential cluster and the minimum density that is required for stars to be assigned to a cluster. The results are plotted in Fig.~\ref{Fig:clustering-dsct}, where the colors represent the label of the cluster to which a star belongs.

The plot shows that the data space is separated into two clusters: one cluster of $295$ stars and one of $589$ stars. We take a random sample of stars from each cluster and perform a visual inspection of their light curves, amplitude spectra and MSE curves. This reveals that cluster 0 mainly consists of hybrid pulsators with frequencies both in the g-mode and p-mode regime, while cluster 1 mainly consists of $\delta$\,Sct stars with pulsations in the p-mode regime only. Given that HDBSCAN sees clusters as regions in the data space with a high density, not every point gets assigned to a cluster. Points in low-density regions that lie further away from cluster centers are in this sense less certain to belong to a particular cluster and HDBSCAN therefore marks these uncertain points as noise. In this case, we get $99$ stars without cluster assignment. A manual extrapolation of the cluster boundaries by means of a line at the intersection of two clusters boundaries would in this case still correctly assign most of these ``noisy'' stars to one of the two clusters, and therefore also improve the overall classification performance. In Fig.~\ref{Fig:spectra-bowman}, we show the light curve, amplitude spectrum and MSE curve for a representative sample star of each of the clusters from Fig.~\ref{Fig:clustering-dsct}. For the stars that have not been assigned to a cluster, we plot both a star that lies around cluster 1 and a star that lies around cluster 0. One can see that the light curves and amplitude spectra of the stars assigned to cluster 1 are characterized by a high density of p-mode oscillations whereas stars in the ``outskirts'' of the cluster are characterized by a significantly lower number of the detected p modes (with the majority of them also showing low amplitudes). Similarly, stars assigned to cluster 0 exhibit rich spectra of both g- and p-modes whereas stars in the outskirts of the cluster show predominantly one type of modes with the other type being significantly underrepresented.

We validate our results by cross-matching the two discovered clusters with the pulsator types assigned by \citet{Bowman2016}. The results are displayed in Table~\ref{Tab:confusion-dsct}, where the percentages are expressed in terms of the column totals. We achieve a True Positive Rate of $82.5\%$ for the $\delta$\,Sct stars and $78.1\%$ for the hybrid stars, which confirms that the MSE in combination with UMAP and HDBSCAN can successfully separate hybrid pulsators from pure $\delta$\,Sct stars. The labels from \citet{Bowman2016} were added to Fig.~\ref{Fig:clustering-dsct} by means of different marker shapes. These labels were not used during the actual clustering procedure. They were only added afterward for the purpose of validation and visualization. We additionally also checked the position in the UMAP plot of two high-amplitude $\delta$\,Sct stars (HADS; see \citealt{McNamara2000} for more details) that were included in the catalog. They appeared to be in the region around the $\delta$\,Sct cluster, but rather in the low-density noisy part and not in the cluster center. The sample size is however too small to draw any firm conclusions from this.

\begin{table}
    \centering
    \makegapedcells
    \begin{tabular}{cc | cc | c}
    \multicolumn{2}{c}{}    &   \multicolumn{2}{c}{\makecell{\textbf{Visual inspection} \\ \textbf{from Bowman et al. (2016)}}} \\
        &       &   $\delta$\,Sct &   hybrid              \\ 
        \cline{2-5}
        \multirow{4}{*}{\rotatebox[origin=c]{90}{\textbf{Cluster}}}
        & 0     & 34 \small{\textit{(5.2\%)}}   & 261 \small{\textit{(78.1\%)}}  & 295             \\
        & 1     &  538 \small{\textit{(82.9\%)}} & 51 \small{\textit{(15.3\%)}}   & 589                \\ 
        & Not assigned    & 77    & 22    & 99                \\ 
        \cline{2-5}
        &       & 649   & 334   & 983
    \end{tabular}
    \caption{Confusion matrix of the cluster assignments calculated with HDBSCAN and the class labels assigned by \citet{Bowman2016} based on visual inspection. The percentages are expressed in terms of the column total. See text for a definition of the HDBSCAN ``0'' and ``1'' clusters.}
    \label{Tab:confusion-dsct}
\end{table}

\begin{figure*}
\centering
   \includegraphics[width=18cm]{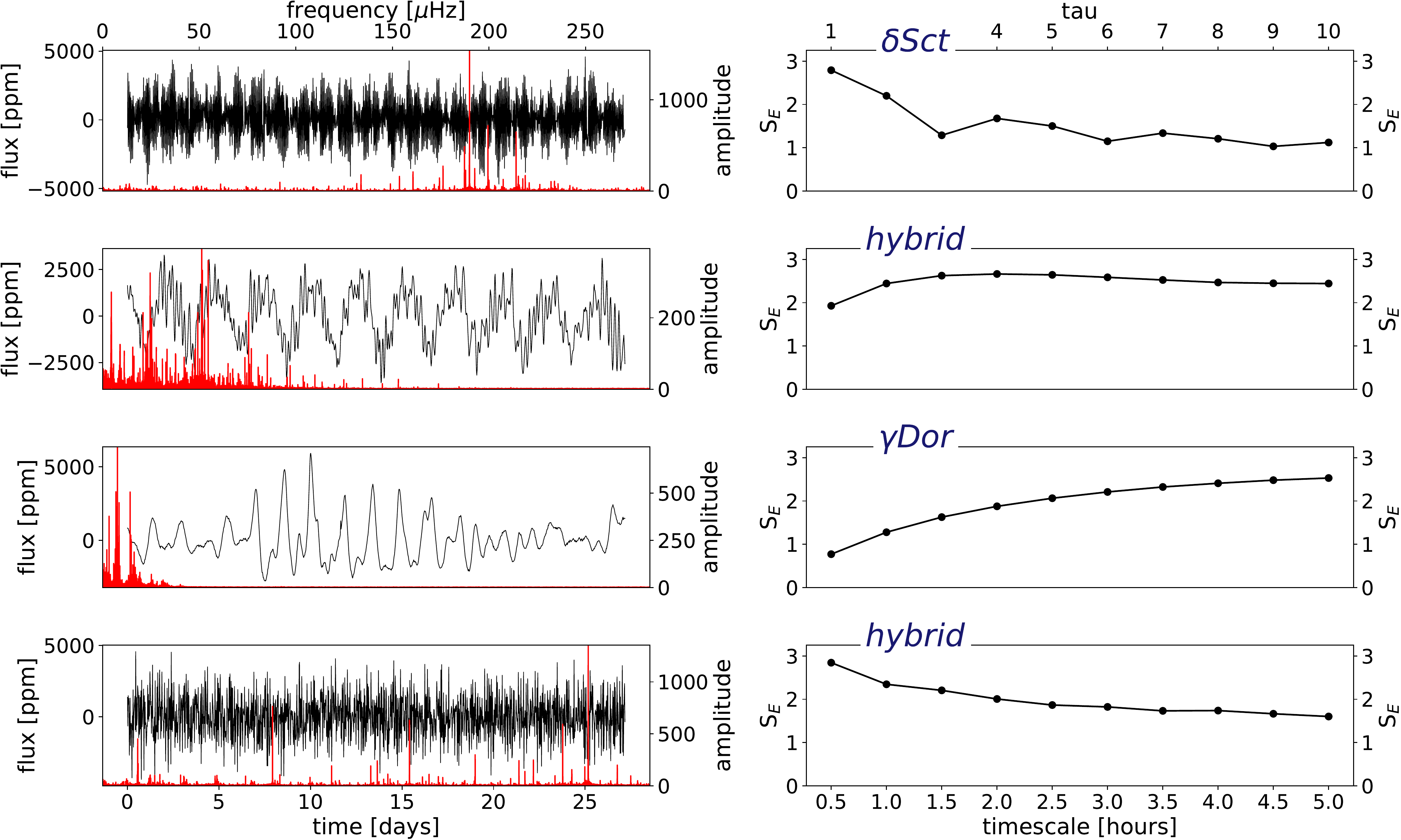}
   \caption{Same as Fig.~\ref{Fig:mse-curves}, but for a $\delta$\,Sct, a $\gamma$\,Dor and two hybrids. The amplitude spectra and multiscale entropy curves were calculated based on 4 yr of data but we only plot a 27 d excerpt of the light curve for clarity.}
    \label{Fig:mse-hybrids}
\end{figure*}

\begin{figure*}
\centering
   \includegraphics[width=18cm]{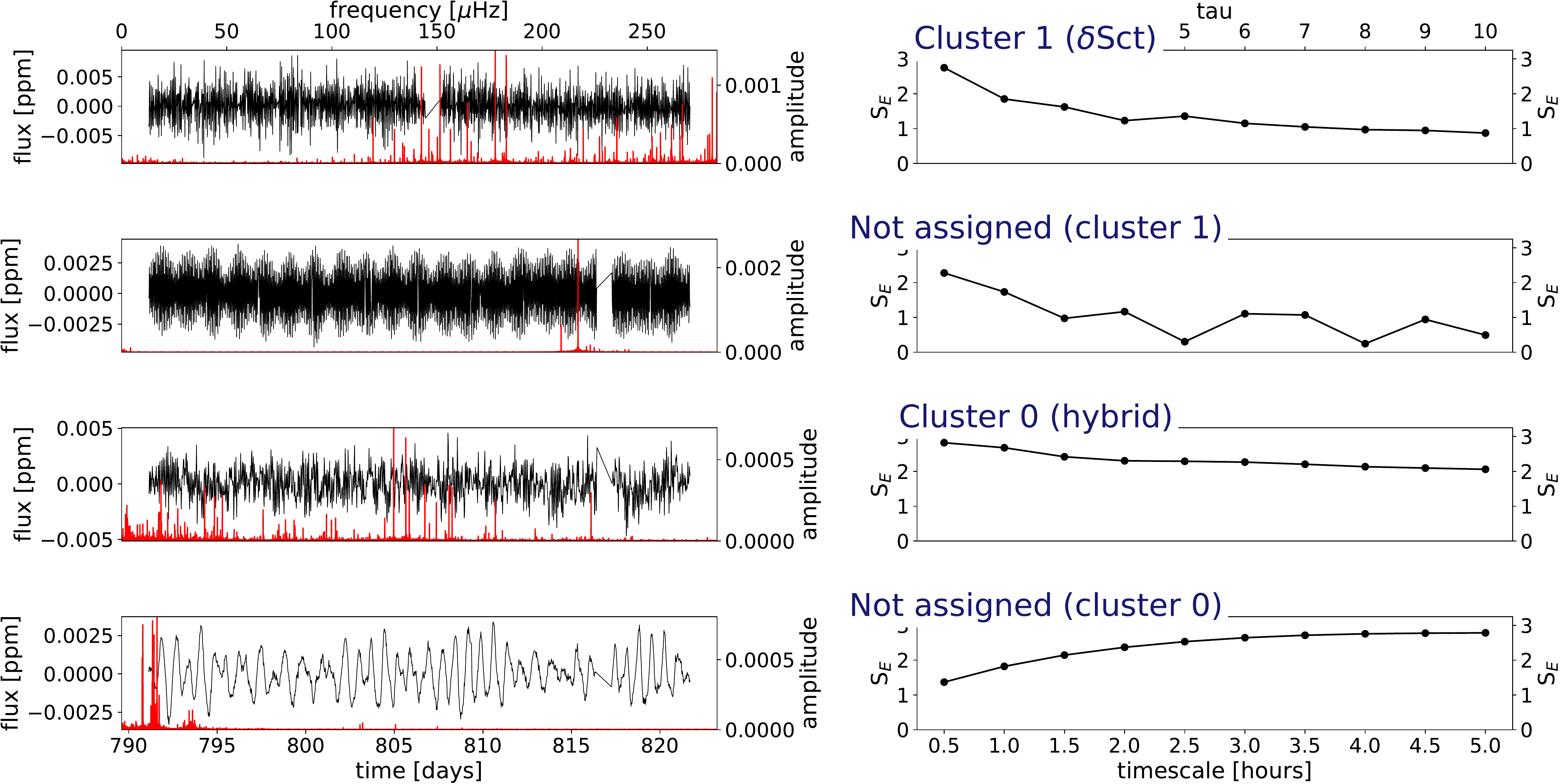}
   \caption{Same as Fig.~\ref{Fig:mse-curves}, but for a sample star drawn from each of the clusters in Fig.~\ref{Fig:clustering-dsct}. The amplitude spectra and multiscale entropy curves were calculated based on 4 yr of data but we only plot a 27 d excerpt of the light curve for clarity.}
    \label{Fig:spectra-bowman}
\end{figure*}

\subsection{\texorpdfstring{$\gamma$}{gamma} Dor star catalog}

\begin{figure}
\centering
   \includegraphics[width=9cm]{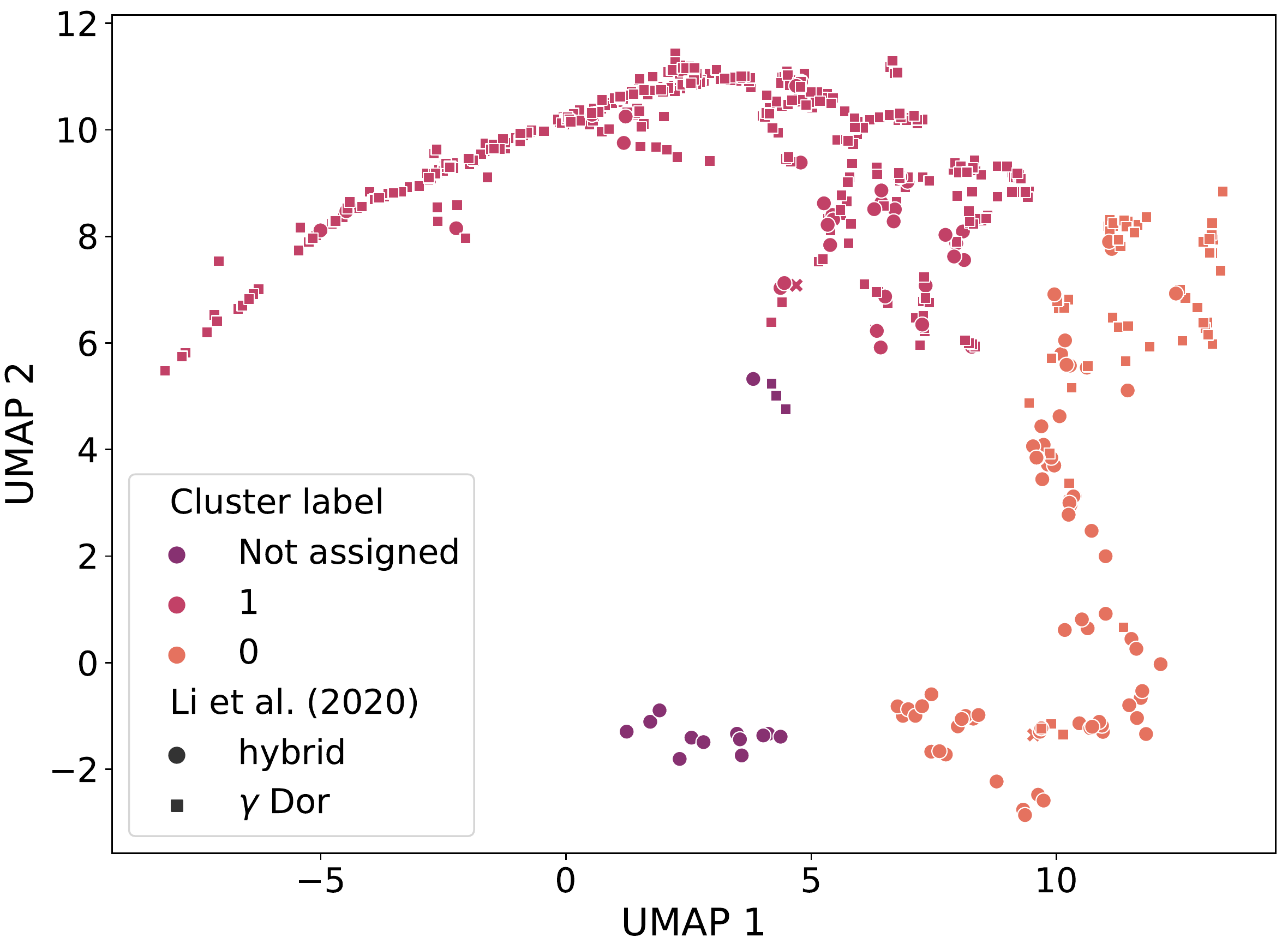}
   \caption{Same as Fig.~\ref{Fig:clustering-dsct}, but for the $\gamma$\,Dor sample from \citet{Li2020}. The corresponding confusion matrix is shown in Table~\ref{Tab:confusion-gdor}.}
    \label{Fig:clustering-gdor}
\end{figure}

We take the catalog of 611 $\gamma$\,Dor stars compiled by \citet{Li2020} and use long-cadence \textit{Kepler} PDC\_SAP data to create our light curves. We stitched the individual quarters together by first detrending the light curve of each quarter with a 2nd- (Q1-Q17) or 1st-order polynomial (Q0). We remove 6 stars from the catalog as they are known to be part of a binary system, leaving us with 605 light curves. We then apply the same steps as in Sect.~\ref{Subsect:dsct}, but instead of differentiating hybrid pulsators from pure $\delta$\,Sct stars with only p-modes, we now aim to separate hybrid pulsators from pure $\gamma$\,Dor stars with g-mode pulsations only. We use \verb"dens_lambda" = 2.0 for UMAP due to a different structure of the clusters. The results are plotted in Fig.~\ref{Fig:clustering-gdor} and listed in Table~\ref{Tab:confusion-gdor}.

We find again that UMAP is able to successfully reduce the data to two dimensions, but that in this case the clusters are not as dense as in the $\delta$~Sct star case. The stars with a hybrid pulsation structure are still largely separated from those with g-mode pulsations only, but they are much more spread out in the plot. We find that cluster 0 contains the hybrid pulsators and cluster 1 the pure $\gamma$\,Dor stars, with respectively $59$ and $426$ stars in each cluster. We show a prototypical example of a light curve, amplitude spectrum and MSE for each type in, respectively, the fourth and third row of Fig.\,\ref{Fig:mse-hybrids}.

\begin{figure}
\centering
   \includegraphics[width=9cm]{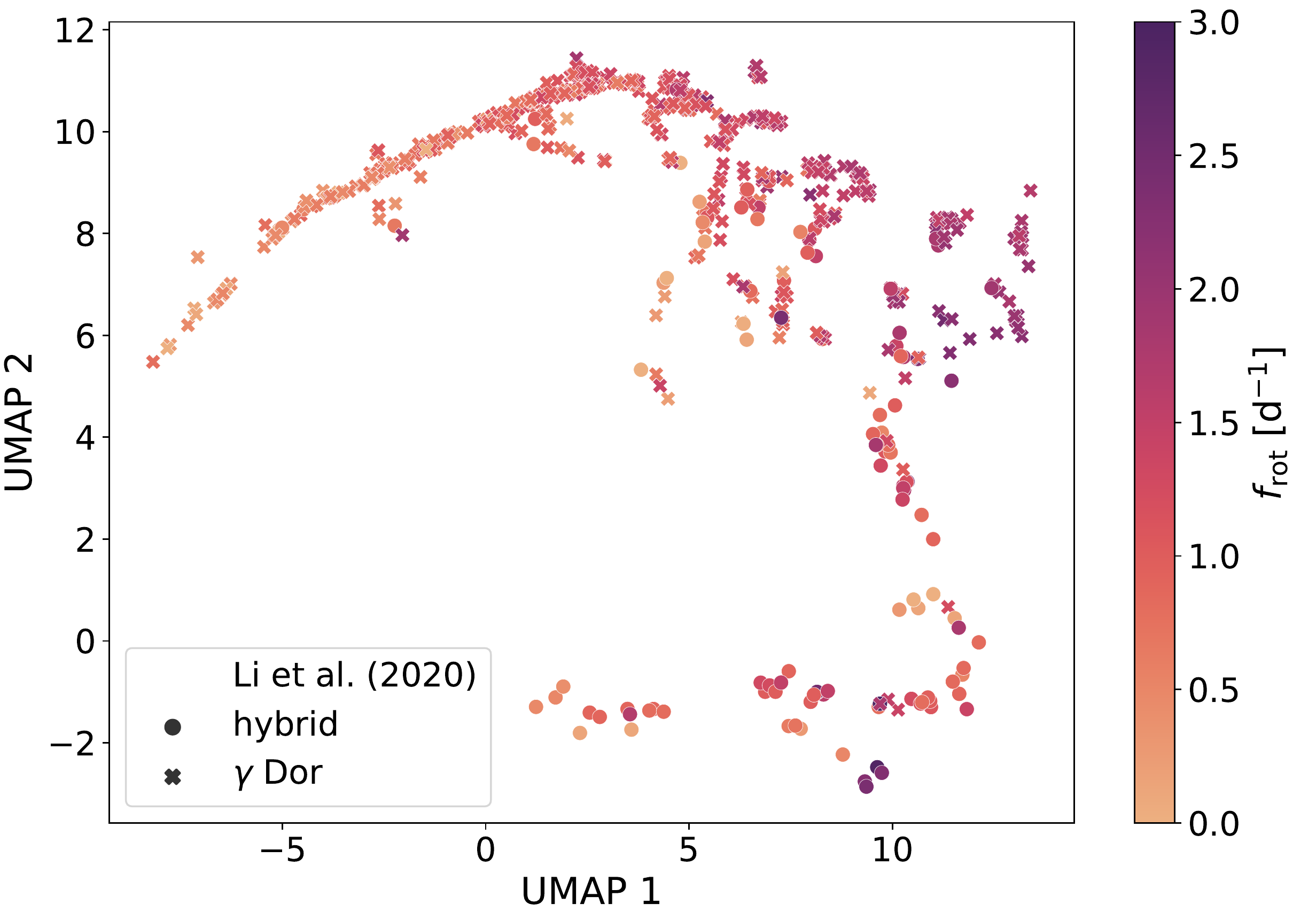}
   \caption{Same as Fig.~\ref{Fig:clustering-gdor}, but color-coded according to the near-core rotation rates from \citet{Li2020}. The figure shows that the structure of the UMAP values, and thus the MSE, can partially be explained by $f_{\rm rot}$ }
    \label{Fig:frot-gdor}
\end{figure}

\begin{figure*}
\centering
   \includegraphics[width=18cm]{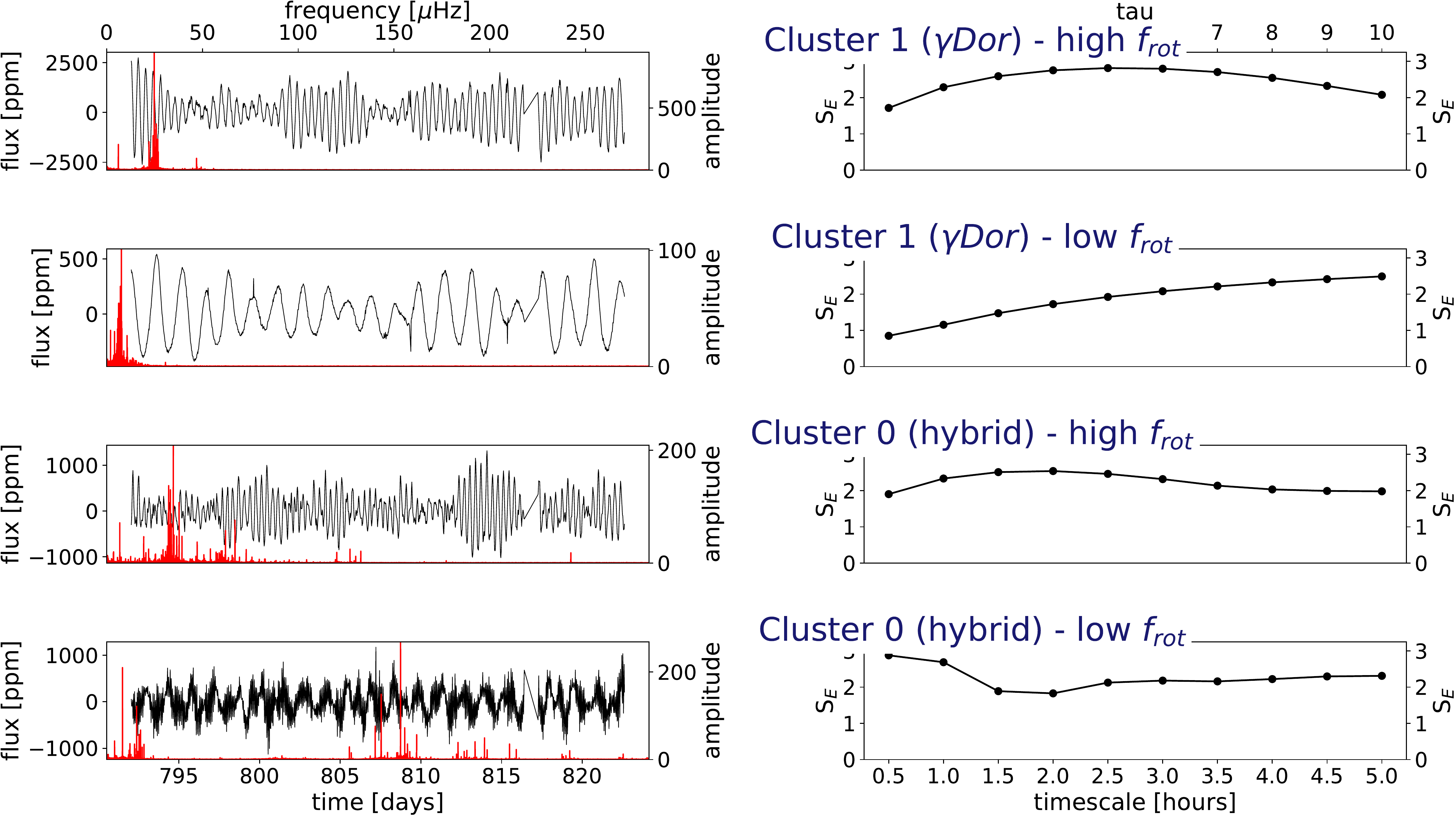}
   \caption{Same as Fig.~\ref{Fig:mse-curves}, but for a sample star drawn from each of the clusters in Fig.~\ref{Fig:clustering-gdor} according to the $f_{rot}$ values in Fig.~\ref{Fig:frot-gdor}. The amplitude spectra and multiscale entropy curves were calculated based on 4 yr of data but we only plot a 27 d excerpt of the light curve for clarity.}
    \label{Fig:spectra-li}
\end{figure*}

One possible cause for the larger dispersion of the data points in Fig.~\ref{Fig:clustering-gdor} is the effect of rotation. Rotation globally shifts the frequencies of g-modes to higher levels and can therefore potentially push, otherwise similar, light curves away from one another \citep{VanReeth2015,Papics2017,Szewczuk2021}. This results in a larger variety of MSE curves and thus in a data space with a lower density, especially if the data set is small. We show the effect of rotation on the MSE in Fig.~\ref{Fig:frot-gdor}, where, as in Fig.~\ref{Fig:clustering-gdor}, the two UMAP components of the MSE curves are plotted, but this time the stars are color-coded according to their near-core rotation rates $f_{\rm rot}$, as determined by \citet{Li2020}. This clearly shows a strong correlation between the elongated and more dispersed shape of the plot and $f_{\rm rot}$. We also take two representative sample stars from each of the two clusters in Fig.~\ref{Fig:clustering-gdor}, one with a high and one with a low value for $f_{rot}$, to illustrate the differences in frequency values and MSE curves that occur due to rotation. The plots are shown in Fig.~\ref{Fig:spectra-li}, where one clearly recognizes a shift of g-mode frequencies toward the higher-frequency regime in high $f_{\rm rot}$ stars. Training a random forest regressor with the MSE values as input and $f_{\rm rot}$ as output, results in $R^2 \approx 0.50 \pm 0.06$, confirming their relation. The variation in $R^2$ depends on the initialization parameters of the random forest and the split of the training and testing set. Fitting the data with a linear regression model gives a similar result, with $R^2 \approx 0.57$. The structure of the plot might also be affected by the fact that the instrumental trends in the light curves are still relatively large compared to the low-amplitude g-modes of the stars.

The more sparsely sampled data space gives HDBSCAN a more difficult time to cluster hybrid pulsators, resulting in a true positive rate of $59.0\%$ for this class, which is significantly lower than in the $\delta$\,Sct star case. Including the unassigned points by drawing a line at the separation point of the two clusters, increases the true positive rate to around $68.0\%$ for the hybrid class. HDBSCAN has fewer problems with the pure $\gamma$\,Dor stars here, because they, in contrast, almost all lie on the same line with a relatively high density. This results in a true positive rate of $87.2\%$ for the $\gamma$\,Dor class.

\begin{table}
    \centering
    \makegapedcells
    \begin{tabular}{cc | cc | c}
    \multicolumn{2}{c}{}    &   \multicolumn{2}{c}{\makecell{\textbf{Visual inspection} \\ \textbf{from Li et al. (2020)}}} \\
        &       &   $\gamma$\,Dor &   hybrid              \\ 
        \cline{2-5}
        \multirow{4}{*}{\rotatebox[origin=c]{90}{\textbf{Cluster}}}
        & 0     & 59 \small{\textit{(12.1\%)}}   & 69 \small{\textit{(59.0\%)}}  & 128             \\
        & 1     &  426 \small{\textit{(87.3\%)}} & 35 \small{\textit{(29.9\%)}}   & 461                \\ 
        & Not assigned    & 3    & 13    & 16                \\ 
        \cline{2-5}
        &       & 488   & 117   & 605
    \end{tabular}
    \caption{Same as Table~\ref{Tab:confusion-dsct} but for the $\gamma$\,Dor-type and hybrid pulsators from \citet{Li2020}.}
    \label{Tab:confusion-gdor}
\end{table}

\section{Conclusions and future prospects}
\label{sect:conclusions}

The MSE is a powerful tool to characterize stellar pulsations based on high-cadence uninterrupted photometric light curves. It provides insights into the structure of the variability, relative amount of short- and long-term variability and is related to the frequency content of a star, with the rotational frequency of $\gamma$\,Dor stars in addition to their pulsations particular. Our analysis reveals that there exists a continuum in which the MSE curves move from $\delta$\,Sct stars toward hybrid stars toward $\gamma$\,Dor stars, illustrating its descriptive power for pulsational variability.

We have leveraged this strength of the MSE to characterize stellar pulsation structure for the development of a clustering tool that can differentiate hybrid pulsators with both p- and g-modes from pure pulsators that, respectively, exhibit only p-mode ($\delta$\,Sct) and g-mode ($\gamma$\,Dor) pulsations. The framework constitutes an important step toward performing a more in-depth classification of stars observed by the \textit{Kepler} and TESS missions. It has the potential to serve as a "Level 2" classifier in the T'DA classification framework from \citet{Audenaert2021}, in which high-level variability classes such as $\delta$\,Sct and $\gamma$\,Dor stars are subdivided into their more detailed constituents, which in our case are stars with only p- or only g-mode pulsations and stars with both p- and g-mode pulsations simultaneously. Hybrid pulsators are important to the asteroseismic community as they allow for more detailed analyses of the internal core and envelope physics. In particular, hybrids allow us to deduce stellar rotation profiles \citep[][Fig. 4]{Aerts2019} and gain better insights into the mechanisms that drive stellar pulsations. The MSE in itself is also useful for the study of the rotation as it is strongly correlated to the near-core rotation rates of g-mode pulsators. The MSE might allow us to estimate these near-core rotation rates without having to perform detailed asteroseismic analyses.

The benefit of our MSE-UMAP-HDBSCAN clustering methodology is that, in contrast to supervised classification, it does not require any labeled training samples and the results are highly interpretable given that for the classification we only rely on one feature in the time domain. Operating in the time domain instead of the Fourier domain also does not incur a large loss of information in our case, because the MSE is able to capture the timescales of the stellar variability patterns. The integration of our more detailed unsupervised classifier into the high-level T'DA supervised classifier is in this sense optimal, as the data first gets structured on a high level by the supervised classifier after which the unsupervised classifier can use this information to obtain more detailed insights. 

Future work should investigate other clustering setups and test whether the methodology can differentiate between other types of pulsation modes. Instead of first reducing the multiscale entropy curves with UMAP and then clustering the UMAP components with with HDBSCAN, the curves could also be clustered directly with sequential data or time series clustering methods, such as k-Shape \citep{Paparrizos2015} or COBRAS-TS \citep{VanCraenendonck2018}. It should also be investigated whether our clustering tool can actually distinguish the high-amplitude $\delta$\,Sct stars from more typical p-mode $\delta$\,Sct stars, by means of a larger data sample for the former. Lastly, it would be interesting to see if the methodology is capable of distinguishing mixed modes from p-modes in red giants, pure pulsators from pulsators with rotational modulation and pure pulsators from pulsators in binary systems. The clustering methodology should then be run on the sets of \textit{Kepler} and TESS $\delta$\,Sct and $\gamma$\,Dor stars that will be returned by the T'DA classifier from \citet{Audenaert2021}. Our methodology will also be included in the variability classification pipelines of the PLATO mission.

\begin{acknowledgements}
The authors respectfully thank the anonymous referee for the thorough yet timely reviews of the manuscript; their {\it enthousiasm\/} is an important encouragement for us.

The research leading to these results has received funding from the European Research Council (ERC) under the European Union's Horizon 2020 research and innovation programme (grant agreement N$^\circ$670519: MAMSIE), from the KU~Leuven Research Council (grant C16/18/005: PARADISE), from the Research Foundation Flanders (FWO) under grant agreement G0H5416N (ERC Runner Up Project), as well as from the BELgian federal Science Policy Office (BELSPO) through PRODEX grant PLATO.

JA also gratefully acknowledges funding from the Research Foundation Flanders (FWO) by means of a long stay travel grant with grant agreement no. V401922N.

The resources and services used in this work were provided by the VSC (Flemish Supercomputer Center), funded by the Research Foundation - Flanders (FWO) and the Flemish Government.

This research has made use of NASA’s Astrophysics Data System, as well as the NASA/IPAC Extragalactic Database (NED) which is operated by the Jet Propulsion Laboratory, California Institute of Technology, under contract with the National Aeronautics and Space Administration.

This paper includes data collected by the Kepler mission. Funding for the Kepler and K2 mission was provided by NASA's Science Mission Directorate. The authors acknowledge the efforts of the Kepler Mission team in obtaining the light curve data and data validation products used in this publication. These data were generated by the Kepler Mission science pipeline through the efforts of the Kepler Science Operations Center and Science Office. The Kepler light curves are archived at the Mikulski Archive for Space Telescopes.

The authors would also like to thank C. Aerts and the MAMSIE/PARADISE team at KU Leuven for their valuable comments and feedback.

\end{acknowledgements}

\bibliographystyle{aa} 
\bibliography{mse.bib}

\begin{appendix}

\end{appendix}

\end{document}